\documentclass[prx,aps,twocolumn,preprintnumbers,amsmath,amssymb,superscriptaddress,showpacs]{revtex4-1}

\usepackage{amsmath}
\usepackage[pdftex]{graphicx}
\usepackage{amssymb}
\usepackage{color}
\usepackage[pdftex,bookmarks=true,colorlinks=true,linkcolor=blue,urlcolor=blue,citecolor=blue]{hyperref}

\newcommand{\be}{\begin{equation}}
\newcommand{\ee}{\end{equation}}

\newcommand{\bit}{\begin{itemize}}
\newcommand{\eit}{\end{itemize}}
\newcommand{\bea}{\begin{eqnarray}}
\newcommand{\eea}{\end{eqnarray}}

\usepackage{epsfig}

\begin{document}

\title
{Heisenberg antiferromagnet on the Husimi lattice }

\author{H. J. Liao}
\affiliation
{Institute of Physics, Chinese Academy of Sciences, P.O. Box 603, Beijing
100190, China}

\author{Z. Y. Xie}
\affiliation
{Institute of Physics, Chinese Academy of Sciences, P.O. Box 603, Beijing
100190, China}
\affiliation
{Department of Physics, Renmin University of China, Beijing 100872, China}

\author{J. Chen}
\affiliation
{Institute of Physics, Chinese Academy of Sciences, P.O. Box 603, Beijing
100190, China}

\author{X. J. Han}
\affiliation
{Institute of Physics, Chinese Academy of Sciences, P.O. Box 603, Beijing
100190, China}

\author{H. D. Xie}
\affiliation
{Institute of Physics, Chinese Academy of Sciences, P.O. Box 603, Beijing
100190, China}

\author{B. Normand}
\affiliation
{Department of Physics, Renmin University of China, Beijing 100872, China}

\author{T. Xiang}\email{txiang@iphy.ac.cn}

\affiliation
{Institute of Physics, Chinese Academy of Sciences, P.O. Box 603, Beijing
100190, China}
\affiliation{Collaborative Innovation Center of Quantum Matter, Beijing
100190, China}

\date{\today}

\begin{abstract}
We perform a systematic study of the antiferromagnetic Heisenberg model on
the Husimi lattice using numerical tensor-network methods based on Projected
Entangled Simplex States (PESS). The nature of the ground state varies
strongly with the spin quantum number, $S$. For $S = 1/2$, it is an algebraic
(gapless) quantum spin liquid. For $S = 1$, it is a gapped, non-magnetic state
with spontaneous breaking of triangle symmetry (a trimerized simplex-solid
state). For $S = 2$, it is a simplex-solid state with a spin gap and no
symmetry-breaking; both integer-spin simplex-solid states are characterized
by specific degeneracies in the entanglement spectrum. For $S = 3/2$, and
indeed for all spin values $S \ge 5/2$, the ground states have $120$-degree
antiferromagnetic order. In a finite magnetic field, we find that, irrespective
of the value of $S$, there is always a plateau in the magnetization at $m =
1/3$.
\end{abstract}


\maketitle

\section{Introduction}

The investigation of low-dimensional quantum antiferromagnets has long been
one of the most active frontiers in condensed matter physics. One of the
most remarkable advances in the understanding of one-dimensional quantum spin
systems is Haldane's conjecture \cite{Haldane_PLA83,Haldane_PRL83}, that
quantum Berry-phase effects cause the low-energy behavior of Heisenberg chains
to depend strongly on the parity of $2S$. Half-odd-integer spin chains have
gapless excitations and power-law decay of their spin correlation functions,
whereas integer-spin chains have a finite excitation gap (the ``Haldane gap'')
and exponentially decaying spin correlations. In fact Lieb, Schultz, and Mattis
were first to prove that the excitation gap for half-odd-integer spin chains
is bounded only by the system size ($\Delta \propto 1/L$) \cite{LSM61,AL86}.

The Haldane conjecture has inspired extensive theoretical studies, especially
on integer-spin systems. Affleck, Kennedy, Lieb, and Tasaki (AKLT) provided
the first rigorous example of a model with a unique ground state, a gap, and
exponentially decaying spin correlation functions \cite{AKLT87,AKLT88}. It
was later found that AKLT states exhibit several exotic features, such as
a nonlocal ``string order'' and edge states, which are properties of all
states within the same Haldane phase \cite{NR89}. A theoretical framework
for understanding these topological properties has been developed recently
\cite{BDGA08,GW09} and used to classify all one-dimensional gapped systems
\cite{CGW_PRB83,CGW_PRB84,SPC11,CGLW13}.

Extensions of the understanding brought by the Haldane conjecture have long
been sought for quantum spin systems in all dimensions higher than 1. The
Lieb-Schultz-Mattis theorem was extended to higher dimensions by Hastings
\cite{Hastings04}. The AKLT construction can be extended to all higher
dimensions in the form of simplex-solid states \cite{Arovas08}, where the
two-site $S = 1/2$ bond singlet of the AKLT state is generalized to an
$N$-site simplex singlet. As for the AKLT states, the simplex-solid state
is an exact ground state of a many-body Hamiltonian, usually with a gap to
all low-energy excitations. The wave function of a simplex-solid state can
be represented by the Projected Entangled Simplex States (PESS) \cite{XCY+14},
providing the foundation for the numerical tensor-network technique we employ
here. However, while the integer-spin case appears to have a number of
higher-dimensional analogs, it has remained unclear whether any quantum
spin system in dimension $d > 1$ can be simultaneously gapless, non-magnetic,
and not break any other symmetries (particularly translational).

In this context, innumerable studies have been performed of highly frustrated
models in two dimensions, including the triangular, Shastry-Sutherland,
$J_1$-$J_2$ square, checkerboard, $J_1$-$J_2$-$J_3$ honeycomb, and other
geometries, as well as of the pyrochlore lattice in three dimensions. However,
the most challenging and enigmatic frustrated system of all has turned out to
be the nearest-neighbor $S = 1/2$ Heisenberg model on the (two-dimensional)
kagome lattice, due to the strong intrinsic frustration of this geometry.
Despite extensive analytical and numerical efforts for almost three decades,
the nature of the ground state and the existence of a spin gap remain as open
questions, with primary candidates including several types of valence-bond
crystal \cite{VBC_kagome}, different gapped Z$_2$ spin liquids
\cite{Z2SL_kagome}, and a gapless, algebraic quantum spin liquid
\cite{U1SL_kagome}. Recently, and in part with a view to solving this
conundrum, more attention has also been paid to kagome Heisenberg
antiferromagnets with higher spins \cite{NS15,PZOP15}. Various proposals
have been put forward for the spin-1 case, including the hexagonal singlet
solid state \cite{Hida00}, the resonating AKLT loop state \cite{YFQ10,LYC+14},
and the trimerized simplex-solid state \cite{CCW09,LLW+15,CL15}, among which
the last has the best variational energy \cite{LLW+15,CL15}. For the $S = 2$
case, a coupled-cluster calculation suggested that the ground state has
$\sqrt 3 \times \sqrt 3 $ antiferromagnetic order \cite{GFB+11}, whereas
the infinite Projected Entangled Pair States (iPEPS) algorithm indicates
a (topologically trivial) spin liquid with a spin gap and no symmetry
breaking \cite{PZOP15}.

The Husimi lattice \cite{Husimi50}, shown in Fig.~\ref{Lattice}(a), is an
infinitely nested set of corner-sharing triangles. Although the local
geometry is identical to that of the kagome lattice [Fig.~\ref{Lattice}(b)],
the Husimi lattice has weaker geometrical frustration because of its bisimplex
nature \cite{Henley01} and because the triangles never reconnect, giving it a
tree structure. A major consequence of these features is that the Heisenberg
model defined on the Husimi lattice is significantly easier to calculate than
the kagome case. To be specific, the PESS Ansatz defined on the Husimi lattice
is an infinite tree tensor-network state, which, as we discuss in detail in
Sec.~II, can be computed very efficiently by the simple-update approach
\cite{JWX08,LDX12}. This allows us to perform systematic investigations
within the PESS framework of the physical properties of the ground state
for Heisenberg models of arbitrary spin quantum number, $S$, on the Husimi
lattice. Within the confines of the Husimi geometry, we may thus characterize
the unique quantum ground states at small $S$ and the quantum-to-classical
crossover with increasing $S$. Beyond the Husimi lattice, its geometrical
similarity to the kagome lattice (Fig.~\ref{Lattice}) suggests the possibility
of many similar physical properties \cite{EZ93,HT09}, and thus such an
investigation may shed new light on the nature of the kagome system.

With this motivation, here we study the properties of the antiferromagnetic
Heisenberg model on the Husimi lattice for spin quantum numbers up to $S = 4$,
working directly in the theromdynamic limit by the PESS technique \cite{XCY+14}.
We find a wide variety of quantum ground states at zero field, ranging from a
gapless spin liquid for $S = 1/2$ through different types of gapped,
simplex-solid state for $S = 1$ and $S = 2$, to (120$^0$-)ordered
N\'eel-type antiferromagnets for $S = 3/2$ and $S \ge 5/2$. Despite these
differences, every single model shows a 1/3 plateau in the magnetization,
suggesting a further rich variety of quantum states at finite applied fields.

This paper is organized as follows. In Sec.~II, we give a brief introduction
to the model and to the Husimi lattice, we review the simplex-solid states
and present their generalization to situations with broken translational and
spin symmetries, and we discuss the simple-update method for computations
using the PESS wave function, including of the entanglement spectrum. In
Sec.~III, we present our results for the zero-field energies, spontaneous
magnetizations, and entanglement spectra of Heisenberg models on the Husimi
lattice for spin quantum numbers up to $S = 4$. In Sec.~IV, we extend our
considerations to a finite magnetic field, compute the induced magnetization
curves for all $S$ values, and comment in detail on the state at 1/3 of the
saturation magnetization. Section V contains a discussion and a brief
summary of our results.

\section{MODEL AND METHOD}

\subsection{Model}

We consider the nearest-neighbor antiferromagnetic Heisenberg model in the
presence of an external magnetic field, $h$, applied in the $z$ direction of
spin space, on the Husimi lattice of Fig.~\ref{Lattice}(a). The Hamiltonian
is given by
\be
H = J \sum_{\langle i,j\rangle} {\bf S}_i \cdot {\bf S}_j - h\sum_{i} {\bf S}_{i}^{z},
\label{HMF}
\ee
where ${\bf S}_i$ is the spin-$S$ operator on site $i$, we investigate spin
quantum numbers up to $S = 4$, $\langle i,j \rangle$ denotes the sum over
nearest-neighbor sites, and $J$ is the nearest-neighbor antiferromagnetic
exchange coupling, which is set henceforth as the energy scale ($J = 1$).

\subsection{Properties of the Husimi Lattice}

The Husimi tree, first introduced in statistical mechanics by Husimi
\cite{Husimi50,RU53,HU53}, is a connected graph whose lobes are all
$p$-polygons ($p \ge 2$, where the $2$-polygon is a bond, the $3$-polygon
a triangle, and so on) and whose bonds belong to at most one simple cycle.
If all lobes consist of only one type of $p$-polygon, the system is known as
a pure Husimi tree, of which the simplest is the Cayley tree \cite{Cayley78},
whose lobes consist only of bonds. The Husimi lattice is an infinite pure
Husimi tree. A Husimi lattice can be characterized by two numbers, $p$ and
$z$, where $p$ is the number of edges of the $p$-polygon and $z$ is the
coordination number of each vertex. A Husimi lattice with $z = 4$ and $p > 2$
can be derived from the Bethe lattice \cite{Ostilli12} with coordination
number $p$ if each bond of the Bethe lattice is replaced by a single vertex
and each vertex by a single $p$-polygon.

\begin{figure}[t]
\begin{center}
\epsfig{file=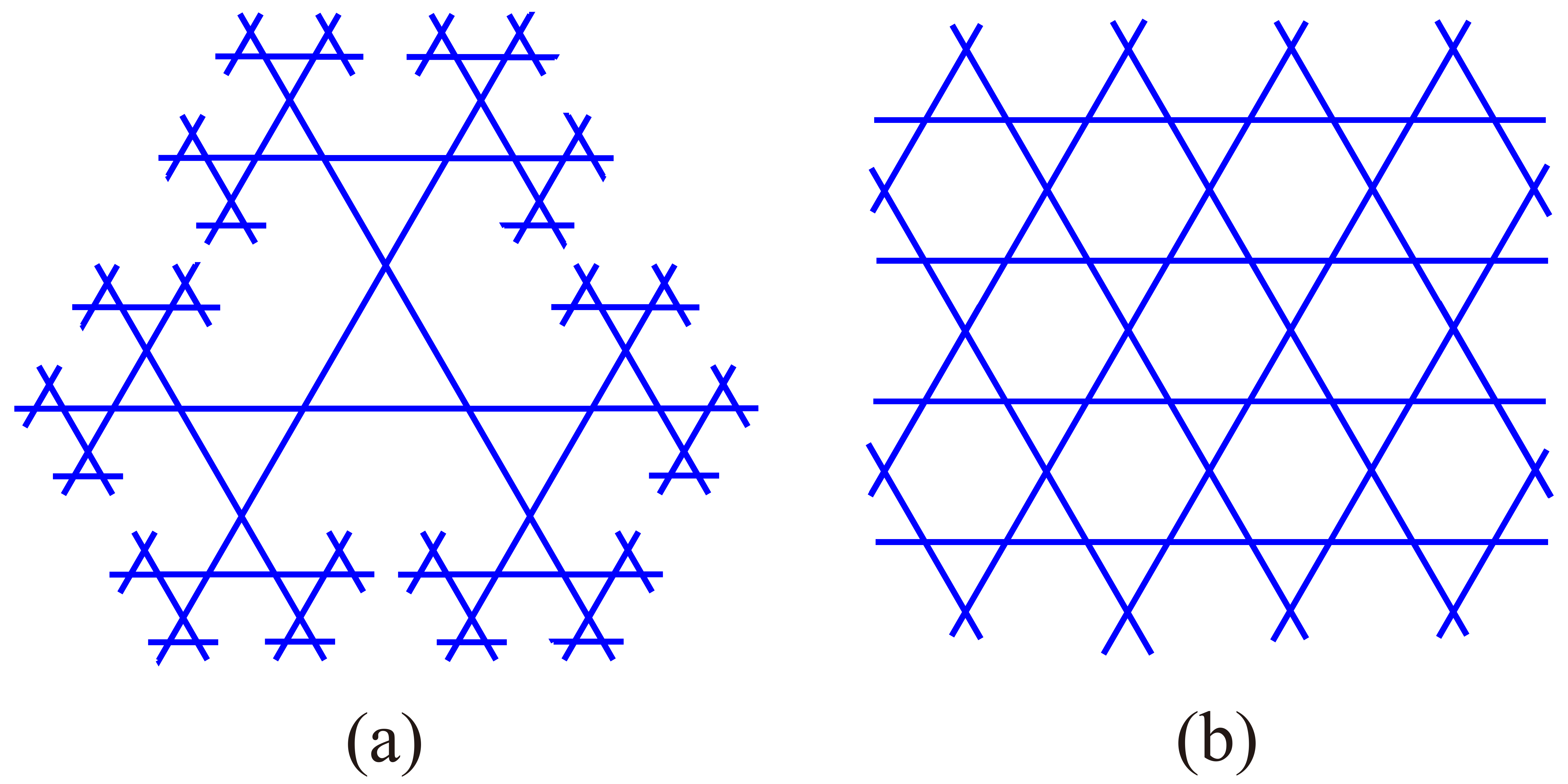,width=8.0cm,angle=0.0}
\end{center}
\caption{(Color online) (a) Triangular Husimi lattice; (b) kagome Lattice.}
\label{Lattice}
\end{figure}

The general quasi-regular tiling $\{{p \atop q}\}$ is composed of two
types of regular polygon with edge numbers $p$ and $q$, which are arranged
alternately around each vertex. The coordination number of all structures
$\{{p \atop q}\}$ is equal to four. The Husimi lattice with $z = 4$ and
$p > 2$ can be also regarded as a limiting case of the quasi-regular tiling
$\{{p \atop q}\}$ in the hyperbolic plane, with $q = \infty$ \cite{MM93}.
The $\{{3 \atop \infty}\}$ and $\{{4 \atop \infty}\}$ Husimi lattices are
also known respectively as the triangular and square Husimi lattices, and in
this sense the system on which we focus here is more accurately specified as
the triangular Husimi lattice [Fig.~\ref{Lattice}(a)]. The kagome lattice
[Fig.~\ref{Lattice}(b)] is the quasi-regular structure $\{{3 \atop 6}\}$
and we stress again that the two share the same local geometry
(Fig.~\ref{Lattice}).

There is, however, an essential difference between the Husimi tree and the
Husimi lattice. In the same way that the Cayley tree differs from the Bethe
lattice \cite{Ostilli12}, the Husimi lattice has no center and no boundary,
and all of its vertices (or polygons) are equivalent, thereby preserving
translational invariance. By contrast, the Husimi tree does have a central
vertex (or polygon) and a boundary; further, the number of vertices (polygons)
grows exponentially with distance from the center, and thus the ratio between
the number of vertices (polygons) on the boundary and that in the bulk does not
approach zero in the thermodynamic limit. The Husimi tree therefore has very
strong finite-size effects and the physical properties of any model defined
on the finite tree may be very different from those of a model on the infinite
lattice. Normally a model defined on the Husimi lattice is much more suitable
to represent a real physical system than a model on the Husimi tree
\cite{Ostilli12}.

\subsection{Simplex-Solid States}
\label{SimplexSS}

Here we review simplex-solid states and PESS. The simplex-solid state of an
SU(N) quantum antiferromagnet was first introduced by Arovas \cite{Arovas08}
and can be regarded as a generalization of the AKLT construction \cite{AKLT87}.
In any simplex-solid state, the bond singlets of the AKLT state are extended
to $N$-site simplex singlets. As with the AKLT state, one may construct the
parent Hamiltonian, for which the simplex solid is the exact ground state, as
a sum of particular local projection operators. Simplex solids typically have
a gap to all excitations and short-ranged correlation functions.

PESS were introduced by Xie $et$ $al.$~\cite{XCY+14}, by generalizing the
concept of simplex-solid states to a numerical Ansatz designed to solve the
$S = 1/2$ kagome Heisenberg antiferromagnet. PESS are also an extension of
PEPS \cite{VC04}, sharing a number of the advantages of the PEPS formulation,
including the ability to satisfy the boundary area law of entanglement entropy
and to represent any state if the bond dimension is sufficiently large. Beyond
the PEPS framework, PESS introduce a new type of entangled simplex tensor,
which captures the $N$-body entanglement of the $N$ virtual particles within
an $N$-site simplex (beyond the 2-body entanglement contained in PEPS), and
it is believed that this feature has an essential role in reproducing the
properties of frustrated systems \cite{XCY+14}.

\begin{figure}[t]
\begin{center}
\epsfig{file=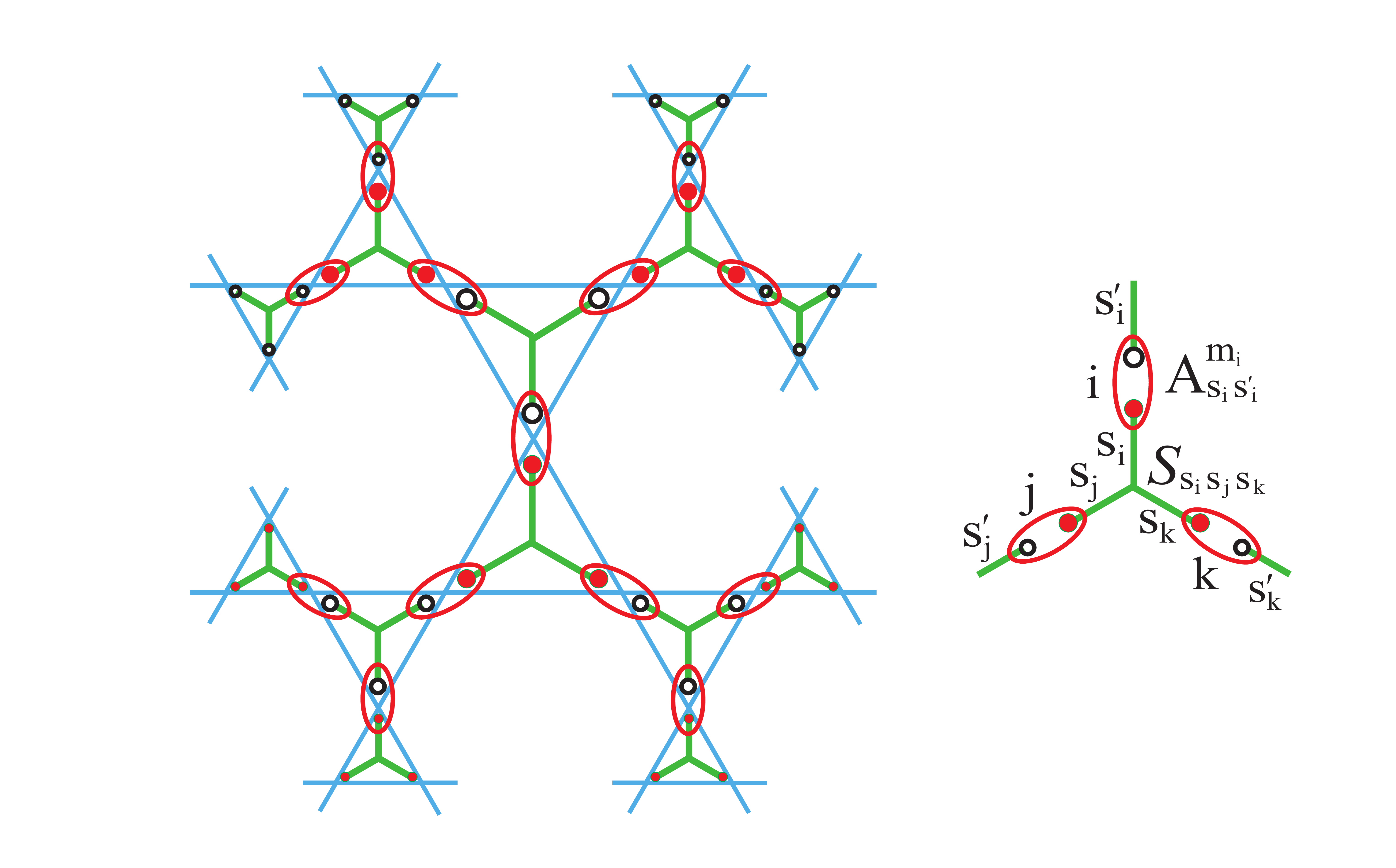,width=8.0cm,angle=0.0}
\end{center}
\caption{(Color online) PESS representation of a simplex-solid state on the
infinite Husimi lattice. Solid circles represent virtual spins $S = n$, open
circles virtual spins $S = n - 1$ ($n$) for a $S = 2n - 1$ ($S = 2n$)
simplex-solid state. Red ellipses represent the projection tensors,
$A_{s_is'_i}^{m_i}$, which project the two virtual spins into the physical spin
subspace. $S_{s_is_js_k}$ denotes the simplex tensor, defined on every triangular
unit.}
\label{SS_state}
\end{figure}

PESS are precisely the tensor-network representation of the simplex-solid
states \cite{XCY+14}. Taking the example of the spin-2 simplex-solid state
on the (triangular) Husimi lattice, the physical $S = 2$ spin at each site
can be treated as a symmetric superposition of two virtual $S = 1$ spins.
Because of the bisimplex property, meaning that two neighboring simplices
share a single site symmetrically, each of the $S = 1$ spins can be assigned
to one of the simplices. Thus each simplex contains three virtual $S = 1$
spins. From the properties of the SU(2) group, decomposition of the product
of three integer spins yields
\begin{eqnarray}
n \otimes n \otimes n = [a_{0}\times0] \oplus \dots \oplus & [ & a_{k} \times
k] \oplus \dots \oplus[a_{3n}\times3n], \nonumber \\
a_{k} = \begin{cases}
2k+1, & k \leq n \\ 3n+1-k, & k > n
\end{cases} & , & \quad k = 0, 1, \dots, 3n,
\label{SU(2)_Integer_Spin}
\end{eqnarray}
where $a_k$ represents the number of times that the $k$th irreducible
representation occurs. $a_0$ is always equal to $1$, i.e.~the three-site
simplex contains a unique spin singlet state. One may thus define a virtual
singlet on each simplex,
\begin{equation}
|\psi_{\alpha} \rangle = \frac{1}{\sqrt{6}} \sum_{\{s\in\alpha\}} S_{s_{i}s_{j}s_{k}}
^{\alpha}| s_{i}, s_{j}, s_{k} \rangle,
\label{Singlet_basis}
\end{equation}
where $s_i$ is a $S = 1$ virtual spin located at site $i$ belonging to the
simplex $\alpha$, and $S_{s_{i}s_{j}s_{k}}^{\alpha}$ is an antisymmetric Levi-Civita
tensor, $\epsilon_{s_{i}s_{j}s_{k}}$. Recovery of the physical spin-2 state
requires the application of a mapping $P_i$ at each site $i$, which projects
the two virtual $S = 1$ spins into the spin-2 subspace,
\begin{equation}
P_{i} = \sum_{s_{i},s_{i}^{\prime}} \sum_{m_{i}} A_{s_{i},s_{i}^{\prime}}^{m_{i}} |m_{i} \rangle
\langle s_{i},s_{i}^{\prime}|,
\end{equation}
where $|m_i \rangle$, with $m_i = 0, \pm 1, \pm 2$, is a basis state of the
physical $S = 2$ spin at site $i$. $A_{s_{i},s_{i}^{\prime}}^{m_{i}}$ is the
Clebsch-Gordan coefficient symmetrizing two virtual $S = 1$ spins and
has nonvanishing components $A_{11}^{2} = A_{33}^{-2} = 1$, $A_{12}^{1} =
A_{21}^{1} = A_{23}^{-1} = A_{32}^{-1} = 1/\sqrt{2}$, $A_{13}^{0} = A_{31}^{0} =
1/\sqrt{6}$, and $A_{22}^{0} = 2/\sqrt{6}$ \cite{XCY+14}. Finally, the
tensor-network representation of this simplex-solid state is the PESS
\begin{eqnarray}
|\Psi \rangle & = & \bigoplus_{i} P_{i} \prod_{\alpha} |\psi_{\alpha} \rangle \\
& = & \mathrm{Tr} (...S_{s_{i}s_{j}s_{k}}^{\alpha} A_{s_{i},s_{i}^{\prime}}^{m_{i}}
A_{s_{j},s_{j}^{\prime}}^{m_{j}} A_{s_{k},s_{k}^{\prime}}^{m_{k}}...) |... m_{i} m_{j}
m_{k} ... \rangle. \nonumber
\label{PESS_wavefunction}
\end{eqnarray}

This description can be extended to any higher even-integer spin. A physical
$S = 2n$ spin is regarded as a symmetric superposition of two virtual $S = n$
spins and the three spins in each simplex are combined to form an SU(2)
simplex singlet. By covering the lattice with equivalent simplex singlets,
one obtains a class of simplex-solid states breaking no lattice symmetries.
Their parent Hamiltonians are readily constructed in terms of local projection
operators. Because half of the virtual spins at the three vertices of any
given simplex are combined into a singlet ($S = 0$), the total spin on each
simplex cannot exceed $S = 3n$. The uniform simplex-solid states are therefore
the exact ground state of the Hamiltonian
\begin{equation}
H = \sum_{m=3n+1}^{6n} J_{m} \sum_{\langle ijk \rangle \in \triangle,\nabla} P_{m} (ijk)
\label{SimplexSS_PHamt1}
\end{equation}
where the second sum $\langle ijk \rangle$ is over all simplices ($\triangle$
and $\nabla$), $J_m$ represents a set of non-negative coupling constants, and
$2n$ ($= S$) is the physical spin quantum number at each site. $P_{m}(ijk)$ is
the operator projecting a state at each simplex onto a state with total spin
$m$, which can be expressed as
\begin{equation}
P_{m}(ijk) = \prod_{l\neq m}^{3S} \frac{(\mathbf{S}_{i} + \mathbf{S}_{j} +
\mathbf{S}_{k})^{2} - l(l+1)}{m(m+1) - l(l+1)},
\label{SPO}
\end{equation}
where $\mathbf{S}_{i}$, $\mathbf{S}_{j}$, and $\mathbf{S}_{k}$ are the vector
spin operators on the three sites of every simplex.

\subsection{PESS States with Broken Symmetry}
\label{SBS}

\subsubsection{Broken Translational Symmetry}
\label{BTS}

To extend the discussion of the previous subsection to systems with arbitrary
odd-integer spin, a physical $S = 2n - 1$ spin must be regarded as a symmetric
superposition of a virtual $S = n$ spin and a virtual $S = n - 1$ spin. One
possible distribution of these unequal spins is to assign the $S = n$ spins
to one type of simplex, for example the upward-oriented triangles ($\Delta$,
referred to henceforth as ``up-triangle''), and the $S = n - 1$ spins to
the other ($\nabla$, henceforth ``down-triangle''). Following
Eq.~(\ref{SU(2)_Integer_Spin}), the three $S = n$ spins on the up-triangles
combine to form an SU(2) spin singlet, and so do the three $S = n - 1$ spins
on the down-triangles. A non-uniform simplex-solid state can be obtained by
arranging the two inequivalent types of simplex singlet in an alternating
pattern on the Husimi lattice, with the inequivalence causing a two-fold
ground-state degeneracy. This class of simplex-solid states breaks lattice
inversion symmetry and favors trimerization \cite{CCW09}. Because the total
spin of the three sites on each bond pair $\langle$, spanning every pair of
inequivalent simplices, cannot exceed $2S$, these simplex-solid states are
exact zero-energy eigenstates of the parent Hamiltonian \cite{Arovas08}
\begin{equation}
H = \sum_{m=2S+1}^{3S} J_{m} \sum_{\langle ijk \rangle \in \langle} P_{m}(ijk),
\label{SimplexSS_PHamt2}
\end{equation}
where the second sum $\langle ijk \rangle$ is over all three-site trios
$(ijk)$ defining one bond on each type of simplex, $J_m$ is a set of
non-negative coupling constants, $S$ is the physical (site) spin quantum
number, and $P_{m}(ijk)$ is the spin projection operator defined by
Eq.~(\ref{SPO}), but with the three sites specified to be on the same
trio $\langle$.

The simplest example of a non-uniform simplex-solid state is obtained for
$S = 1$. This is readily expressed as a PESS wavefunction with bond dimension
$D = 4$, where the two virtual spins at every lattice site each have four basis
states, $|1 \rangle \equiv |\!\! \uparrow \rangle$, $|2 \rangle \equiv |0
\rangle$, $|3 \rangle \equiv |\!\! \downarrow \rangle$, and $|4 \rangle
\equiv |\emptyset \rangle$, representing respectively the components of a
spin-1 triplet and a spin-0 net singlet. The nonvanishing components of the
two simplex tensors are $S_{ijk}^{\triangle} = \frac{1}{\sqrt{6}} \epsilon_{ijk}$
$(i,j,k = 1,2,3)$ and $S_{444}^{\nabla} = 1$, and of the projection tensor are
$A_{14}^{+1} = A_{24}^{0} = A_{34}^{-1} = A_{41}^{+1} = A_{42}^{0} = A_{43}^{-1} = 1,
A_{12}^{+1} = A_{13}^{0} = A_{23}^{-1} = 1/\sqrt{2}$, and $A_{21}^{+1} = A_{31}^{0}
 = A_{32}^{-1} = -1/\sqrt{2}$; the parent Hamiltonian is given simply by
\begin{equation}
H = \sum_{\langle ijk \rangle \in \langle} P_{3}(ijk).
\label{SimplexSS_PHamtS=1}
\end{equation}

We comment that there is no corresponding simplex-solid state on the triangular
Husimi or kagome lattice for the case when the spin quantum number is
half-odd-integer. In this situation, and indeed for any other with an odd-site
simplex, decomposing the product of an odd number of half-odd-integer spins
cannot yield a total spin singlet.

\begin{figure}[t]
\begin{center}
\epsfig{file=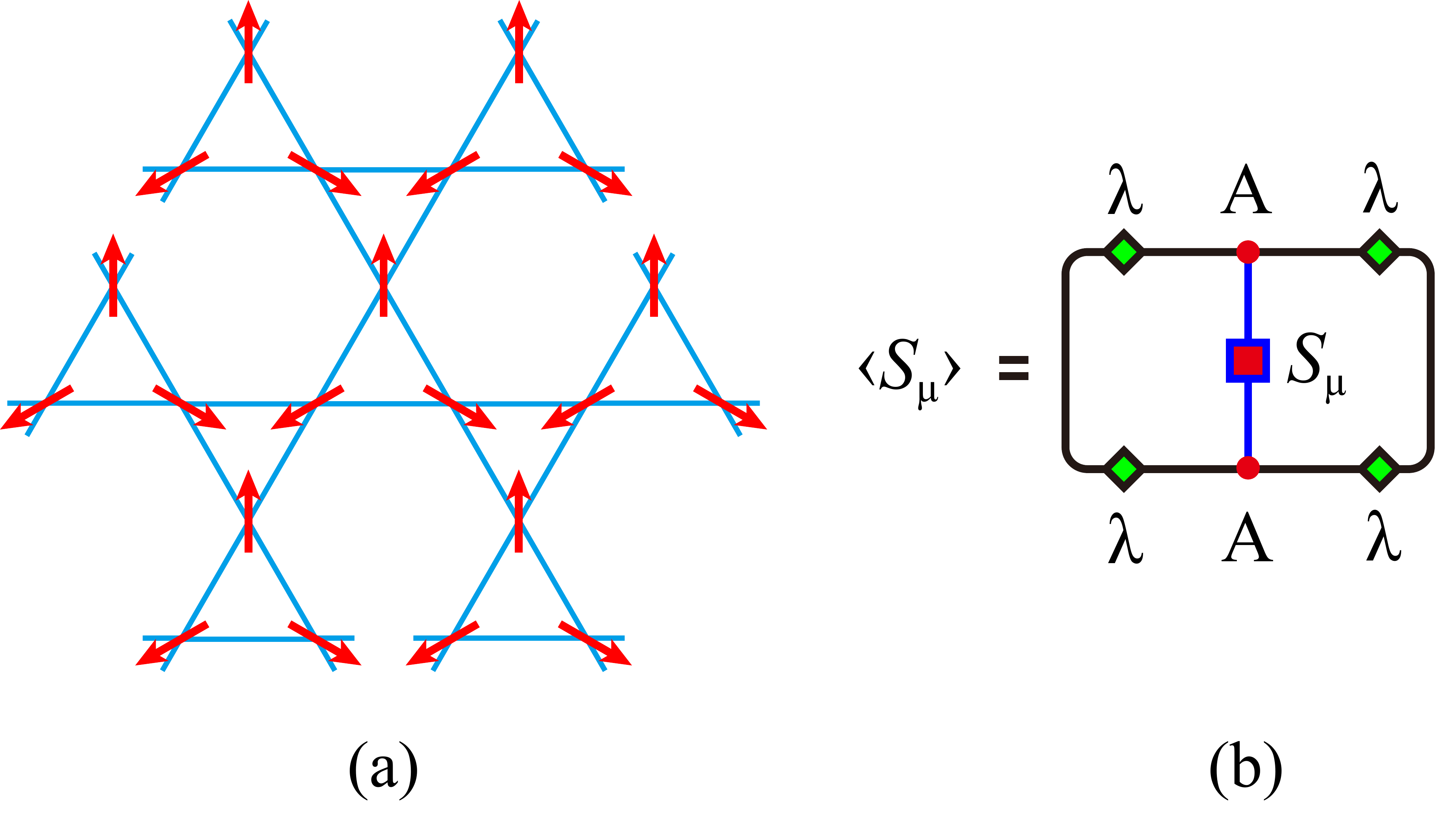,width=8.0cm,angle=0.0}
\end{center}
\caption{(Color online) (a) Schematic representation of the antiferromagnetic
state with 120-degree N\'eel order on the Husimi lattice, where the arrows
denote the orientations of the spins. (b) Graphical representation of the
expectation value of the local spin operator, $S_{\mu} (\mu = x,y,z)$,
for a PESS wave function, where $\lambda$ and $A$ are respectively the bond
vector and the projection tensor in the canonical PESS wave function.
Connecting lines denote the contraction of two neighboring tensors.}
\label{AF_120}
\end{figure}

\subsubsection{Spontaneously Broken Spin Symmetry}
\label{BSS}

In two and higher dimensions, and in particular as the spin quantum number
increases, the ground state of any magnetic system usually favors some
type of ordered state. For strongly frustrated antiferromagnets, where
collinear order is precluded, this order tends to be coplanar in the absence
of an external field, and in triangle-based geometries its most common form
is the 120-degree N\'eel order shown in Fig.~\ref{AF_120}(a). To detect this
type of magnetic order in our PESS calculations, one could in principle
calculate the expectation values of the spin operators at every site and
then consider their mutual orientations. For the $3$-PESS wave function
used in this work \cite{XCY+14}, we assume complete translational invariance
and thus we need only calculate the expectation values of the spin operators
$\langle S_{i}^{x}\rangle$, $\langle S_{i}^{y}\rangle$, and $\langle S_{i}^{z}
\rangle$ at the three sites within each of the two types of triangle (up and
down). Because the PESS wavefunction defined on the Husimi lattice can be
expressed in canonical form, as shown in next subsection, the expectation
value of the local spin operator is easy to calculate, as represented
graphically in Fig.~\ref{AF_120}(b). If the ground state has
antiferromagnetic order, the (``transverse'') magnetization,
\begin{equation}
M = \frac{1}{N} \sum_{i} \sqrt{\langle S_{i}^{x} \rangle^{2} + \langle S_{i}^{y}
\rangle^{2} + \langle S_{i}^{z} \rangle^{2}},
\label{Magnetization}
\end{equation}
will have a finite value. The spontaneous magnetization $M$ serves as the
order parameter for the detection of antiferromagnetically ordered states.

\subsubsection{Applied Magnetic Field}

In the presence of a finite field, $h$, in Eq.~(\ref{HMF}), spin rotation
symmetry is broken explicitly. In this case it is the longitudinal
magnetization, defined as
\begin{equation}
M_z = \frac{1}{N} \sum_{i}\langle S_{i}^{z} \rangle,
\label{elm}
\end{equation}
which takes on a finite value. In contrast to numerical approaches implemented
on systems of finite size, where it is necessary to target sectors of specific
total-spin quantum number to reproduce the effects of an external field,
tensor-network techniques are already in the thermodynamic limit and will
return a wave function appropriate for the field applied. We calculate the
longitudinal magnetization associated with this wave function for all values
of $h$ and $S$ in Sec.~IV, and interpret the physical content of the resulting
states.

We comment here that the PESS code we use in this paper is real. Thus it is
possible to represent all magnetic states where the spins are coplanar, which
naturally includes all collinear spin states. In the event that the combination
of frustration and applied magnetic field were to produce non-coplanar spin
states, or in field-free systems with extended frustration showing, for
example, the double-spiral ground state, it would be necessary to use complex
code to obtain an accurate representation. In the present case, with
nearest-neighbor coupling only and triangular geometries, it is expected
\cite{rtm} that the moments will lie in the same plane for all applied fields.

\begin{figure}[t]
\begin{center}
\epsfig{file=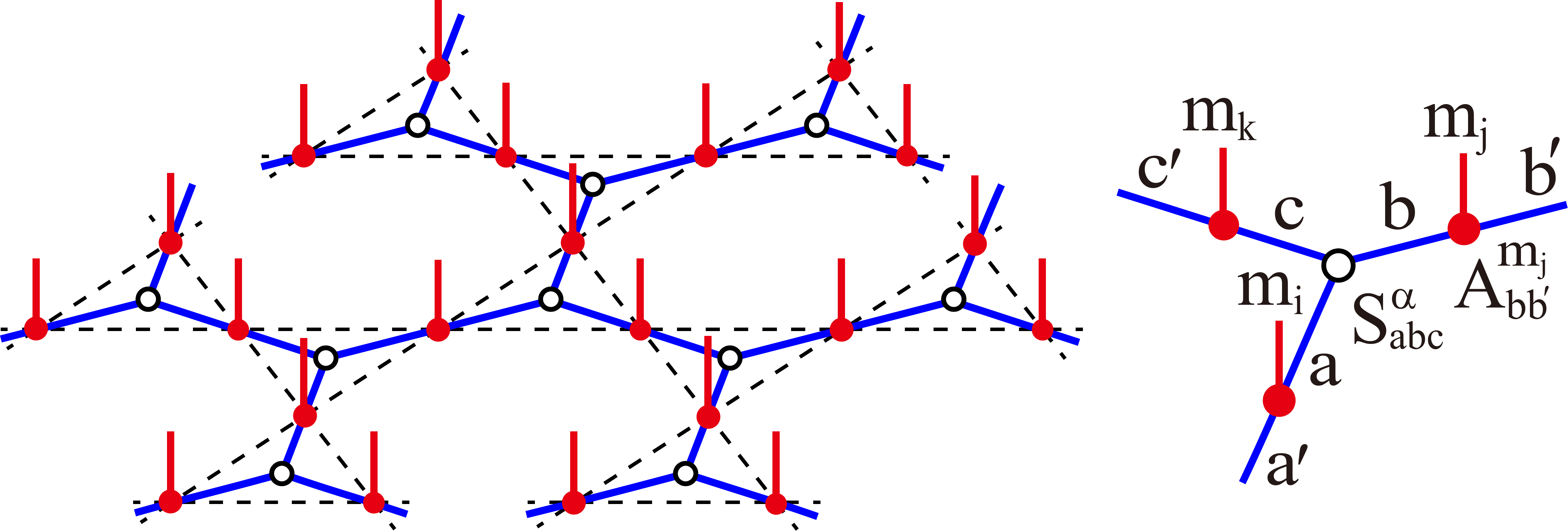,width=8.0cm,angle=0.0}
\end{center}
\caption{(Color online) Graphical representation of a PESS on the infinite
Husimi lattice (dashed lines). Open black circles represent the three-index
simplex tensor, $S^{\alpha}_{abc}$ ($\alpha = \triangle, \nabla$), solid red
circles the three-index projection tensors, $A_{bb'}^{m_j}$, at each
physical lattice site. Blue lines represent the virtual indices of all
tensors, vertical red lines the physical degrees of freedom, $\{m_j\}$,
of each site.}
\label{PESS_Tensor}
\end{figure}

\subsection{Simple-Update Method and Canonical Form}

We now employ the PESS wave function not as an exact description of
simplex-solid states but as a variational Ansatz to capture the ground-state
properties of an arbitrary spin system \cite{XCY+14}. The PESS wave function
on the triangular Husimi lattice is represented graphically in
Fig.~\ref{PESS_Tensor}. Known as a 3-PESS because its simplex
contains three lattice sites, it is composed of simplex entanglement tensors
and projection tensors. The former, $S^{\triangle}_{abc}$ and $S^{\nabla}_{abc}$,
each have three virtual indices and form a Bethe lattice of simplex triangles,
while the latter, $A_{aa'}^m$, $A_{bb'}^m$, and $A_{cc'}^m$, each have one
physical and two virtual indices and are located at the decorating sites
of this Bethe lattice (the original sites of the Husimi lattice). In contrast
to Sec.~IIC, where the tensor indices for the simplex solid were virtual spin
indices, now $a$, $a'$, $b$, \dots are general virtual indices of the tensor
network and their dimension is the bond dimension, $D$. Each physical index,
$m_i$, runs over the $d = 2S + 1$ physical basis states defined on each
lattice site, $i$.

The ground-state wave function is obtained by repeated application of
imaginary-time evolution operators, $U(\tau) = \exp (-\tau H)$, on an
initial PESS wavefunction, $|\Psi_0 \rangle$, where $\tau$ is taken to be
small. The Hamiltonian is split into
\begin{equation}
H = H_{\triangle} + H_{\nabla},
\end{equation}
where $H_{\triangle}$ and $H_{\nabla}$ contain respectively the Hamiltonian terms
on all up- and down-triangles. Because $H_{\triangle}$ and $H_{\nabla}$ do not
commute, the evolution operator is decomposed approximately into a product
of two near-unitary operators using the Trotter-Suzuki formula,
\begin{equation}
e^{-\tau H} = e^{-\tau H_{\triangle}} e^{-\tau H_{\nabla}} + O(\tau^{2}).
\end{equation}
Each iteration of the projection is then performed in two steps by
successive application of $e^{-\tau H_{\triangle}}$ and $e^{-\tau H_{\nabla}}$ to
the wave function.

Each projection step, or application of $e^{-\tau H_{\alpha}}$ to the wave
function ($\alpha = \triangle,\nabla$), increases the dimensions of the
evolved bonds and thus requires a truncation of the bond dimensions of
the new tensors. During this truncation, it is necessary to consider the
renormalization effect of all the other bonds of the system, which can be
encoded as environment tensors, and in general there are two types of scheme
to simulate their contributions. In the full-update approach, a complete and
accurate environment tensor is calculated at each projection step, but the
rather high computational cost of this process limits the bond dimension
to very small values (approximately $D \leq 6$). A more efficient approach,
the simple-update method \cite{JWX08,XJC+09,ZXC+10,XCY+14}, approximates
the effect of the environment tensor using specific positive bond vectors.
This method turns out to be almost exact for one-dimensional systems
\cite{Vidal0304} and, of key importance for the present study, for systems
defined on the Bethe lattice \cite{LDX12}, as long as the imaginary time
step $\tau$ is taken to be sufficiently small.

The reason for this result lies in the bipartite nature of tensor-network
states defined on open chains and on the Bethe lattice. This allows them to
be written in canonical form (below) and divided into two subsystems under
Schmidt decomposition \cite{DP13}, such that the square of the bond vector
defined on each bond is precisely the eigenvalue of the reduced density
matrix if the tensor-network state is kept in its canonical form
\cite{Vidal0304,LDX12}. The bond vector then contains all entanglement
information between the system and environment subblocks, making the
simple-update method equivalent to the full-update approach for bipartite
tensor-network states. Because PESS defined on the Husimi lattice are also
bipartite tree tensor-network states, one expects that the simple-update
approach will provide an efficient determination of the wave function in
this case.

\begin{figure}[t]
\begin{center}
\epsfig{file=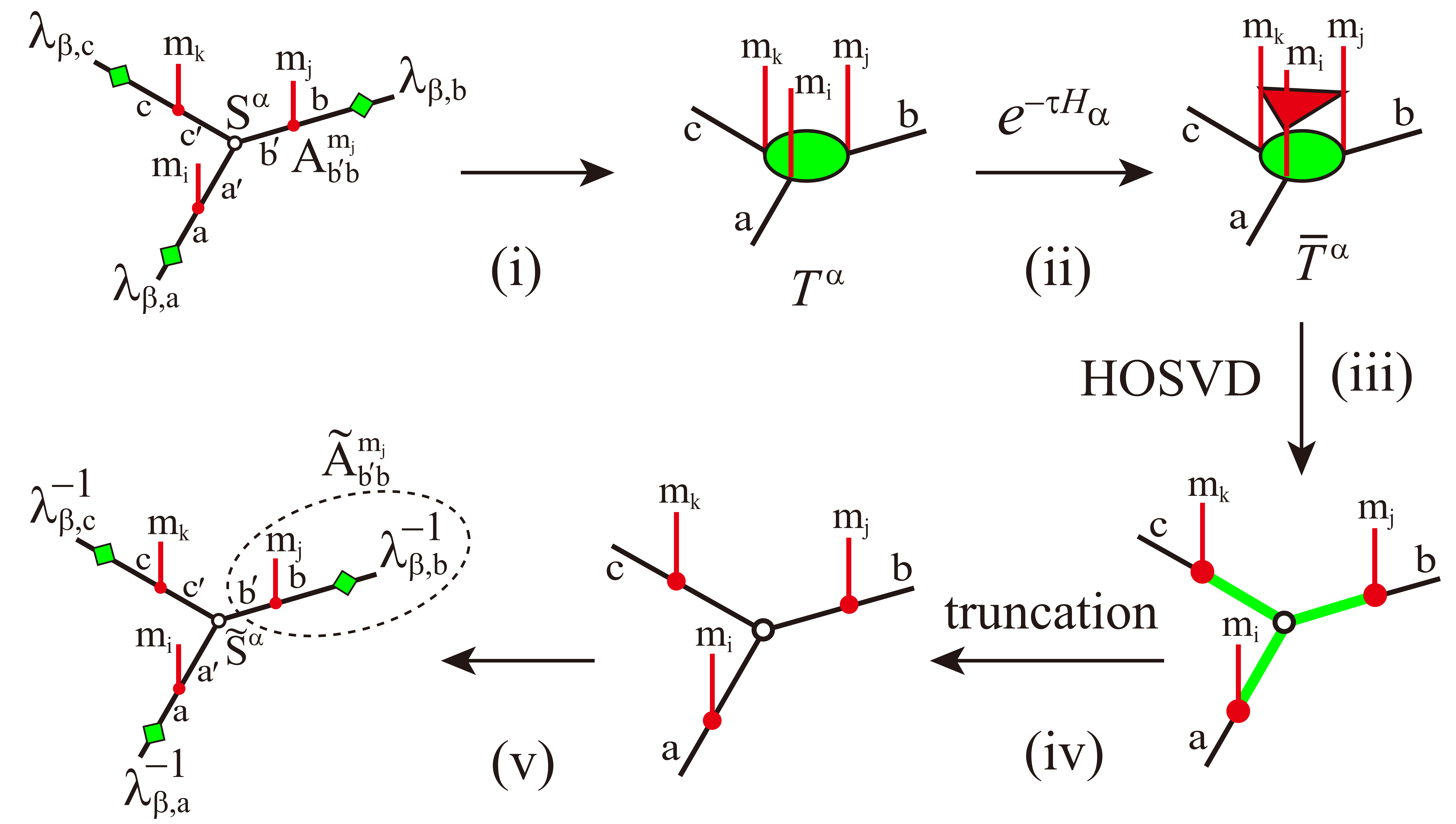,width=8.0cm,angle=0.0}
\end{center}
\caption{(Color online) Graphical representation of the simple-update
procedure for PESS using HOSVD. Details are provided in the text.}
\label{Simple_Update}
\end{figure}

The simple-update procedure for the PESS wave function is specified by the
following steps, represented graphically in Fig.~\ref{Simple_Update}.

(i) Absorb the environment bond vectors $\lambda_{\beta,v}$ into the
projection tensors $A^m_{vv'}$ and then contract the three projection tensors
with the simplex tensor $S^{\alpha}$ to obtain a cluster tensor $T^{\alpha}$,
where $v = {a,b,c}$ represents the virtual bond index, $m$ is the physical
index of the projection tensor, and $\alpha$ and $\beta$ denote two
neighboring simplices (up- or down-triangles).

(ii) Apply the near-unitary evolution operator $e^{-\tau H_{\alpha}}$ to the
cluster tensor $T^{\alpha}$ to obtain a new cluster tensor $\bar{T}^{\alpha}$.

(iii) Decompose the new cluster tensor $\bar{T}^{\alpha}$ into the product
of a renormalized simplex tensor and three renormalized projection tensors
by using higher-order singular value decomposition (HOSVD) \cite{XCQ+12,
XCY+14,LMV2000}. At the same time, one obtains the new bond vectors
$\tilde{\lambda}_{\alpha,v}$.

(iv) Truncate the renormalized simplex tensor and the three renormalized
projection tensors to obtain the updated simplex tensor $\tilde{S}^{\alpha}$
and projection tensors.

(v) Absorb the inverse bond vectors $\lambda_{\beta,v}^{-1}$ into the
truncated projection tensors to obtain the updated projection tensors
$\tilde{A}^m_{vv'}$.

\noindent
One evolution cycle is completed, i.e.~the full tensor network is updated,
by repeating this procedure for each simplex ($\alpha = \triangle,\nabla$).
Repeated update cycles cause the PESS wave function to converge to the ground
state.

The canonical PESS wave function on the Husimi lattice can be obtained by the
simple-update method if the evolution operator is replaced by the identity
operator in step (ii). In the canonical form, all tensors of the PESS should
simultaneously satisfy the canonical conditions
\begin{eqnarray}
& & \sum_{a,b} S_{abc}^{\alpha} (S_{abc^{\prime}}^{\alpha})^{\ast} = \delta_{c,c^{\prime}}
\lambda_{\alpha,c}^{2}, \nonumber \\ & & \sum_{v,m} \lambda_{\alpha,v}^{2}
A_{vv'}^{m} (A_{vv''}^{m})^{\ast} = \delta_{v',v''}, \\ & & \sum_{v,m}
\lambda_{\alpha,v}^{2} A_{v'v}^{m} (A_{v''v}^{m})^{\ast} = \delta_{v',v''}, \nonumber
\label{ecc}
\end{eqnarray}
where $\lambda_{\alpha,v}$ denotes the bond vector for the bond linking the
$S^{\alpha}$ and $A^m_{vv'}$ tensors; the ``left'' and ``right'' conditions
specified in the lower two lines must both be satisfied separately.
Maintaining this canonical form is the key to the success of the
simple-update scheme and is possible due to the bipartite nature of
the system.

\subsection{Entanglement spectrum}
\label{Simplex_ES}

The entanglement spectrum (ES) of a quantum system is defined as the
logarithmic eigenvalue of the reduced density matrix of a many-body
state \cite{LH08}, and provides additional useful information beyond
the entanglement entropy \cite{LH08,LYC+14,DKL14}. As noted above, for
a canonical tensor-network state, the square of the bond vector is
precisely the eigenvalue of the reduced density matrix, and thus it
is easy to calculate the ES using the bond vectors,
\be
\zeta_{\alpha}(i) = -\mathrm{Log_2} \, \lambda_{\alpha,v}^{2}(i), \quad i = 1, 2,
3, \dots, D,
\label{ES}
\ee
where $\lambda_{\alpha,v}$ satisfies the normalization condition $\sum_{i}
\lambda_{\alpha,v}^{2}(i) = 1$.

Taking the $S = 2$ simplex-solid state on the Husimi lattice as an example,
its tensor-network representation is a PESS wave function with bond dimension
$D = 3$, as shown in Subsec.~\ref{SimplexSS}. All the bond vectors obtained
from the canonical form [Eq.~(\ref{ecc})] are $\lambda_{\alpha,v} = (1,1,1)/
\sqrt{3}$, which is the square root of the reduced density matrix associated
with a simplex-singlet state [Eq.~(\ref{Singlet_basis})]. Specifically,
$\hat{\rho}_{s} = tr_{E}(\hat{\rho}) = \sum_{s_{i},s_{i}^{\prime}} (\frac{1}{6}
\sum_{s_{j},s_{k}} \epsilon_{s_{i}s_{j}s_{k}} \epsilon_{s_{i}^{\prime}s_{j}s_{k}})
|s_{i}^{\prime} \rangle \langle s_{i}| = \sum_{s_{i},s_{i}^{\prime}} (\frac{1}{3}
\delta_{s_{i},s_{i}^{\prime}}) |s_{i}^{\prime} \rangle \langle s_{i}|$. This result
states that, if there exists a total singlet on the triangular simplex, then
the ES on the corresponding bond will be three-fold degenerate. For the $S = 2$
simplex-solid state, because there is a total singlet on every simplex, the ES
is three-fold degenerate on every bond of the system. Further, the energy of
this simplex-solid state for the Heisenberg model on the Husimi lattice is
given exactly as $E_0 = -9/2$, and there is no magnetic order.

It is similarly straightforward to obtain the ES of the $S = 1$ simplex-solid
state on the Husimi lattice, whose tensor-network representation is a PESS
wave function with bond dimension $D = 4$, as shown in Subsec.~\ref{SimplexSS}.
For convenience of discussion, henceforth we use the terminology ``A-bond''
(``B-bond'') to denote the bond linking the $S^{\triangle}_{abc}$
($S^{\nabla}_{abc}$) and $A^m_{vv'}$ tensors. After this wave function is
put in canonical form, the bond vectors on the A- and B-bonds are
$(1,1,1,0)/\sqrt{3}$ and $(1,0,0,0)$, indicating that their entanglement
spectra are respectively three- and one-fold degenerate (i.e.~non-degenerate),
as could be expected from the preceding analysis. If the energy of this state
is computed, one indeed finds strong trimerization ``order'' characterized by
$\Delta E = 2|E_{\triangle} - E_{\nabla}|/3 = 2$, where $E_{\triangle}$ ($E_\nabla$)
is the average energy of an up-triangle (down-triangle), but no magnetic order.

\begin{figure}[t]
\begin{center}
\epsfig{file=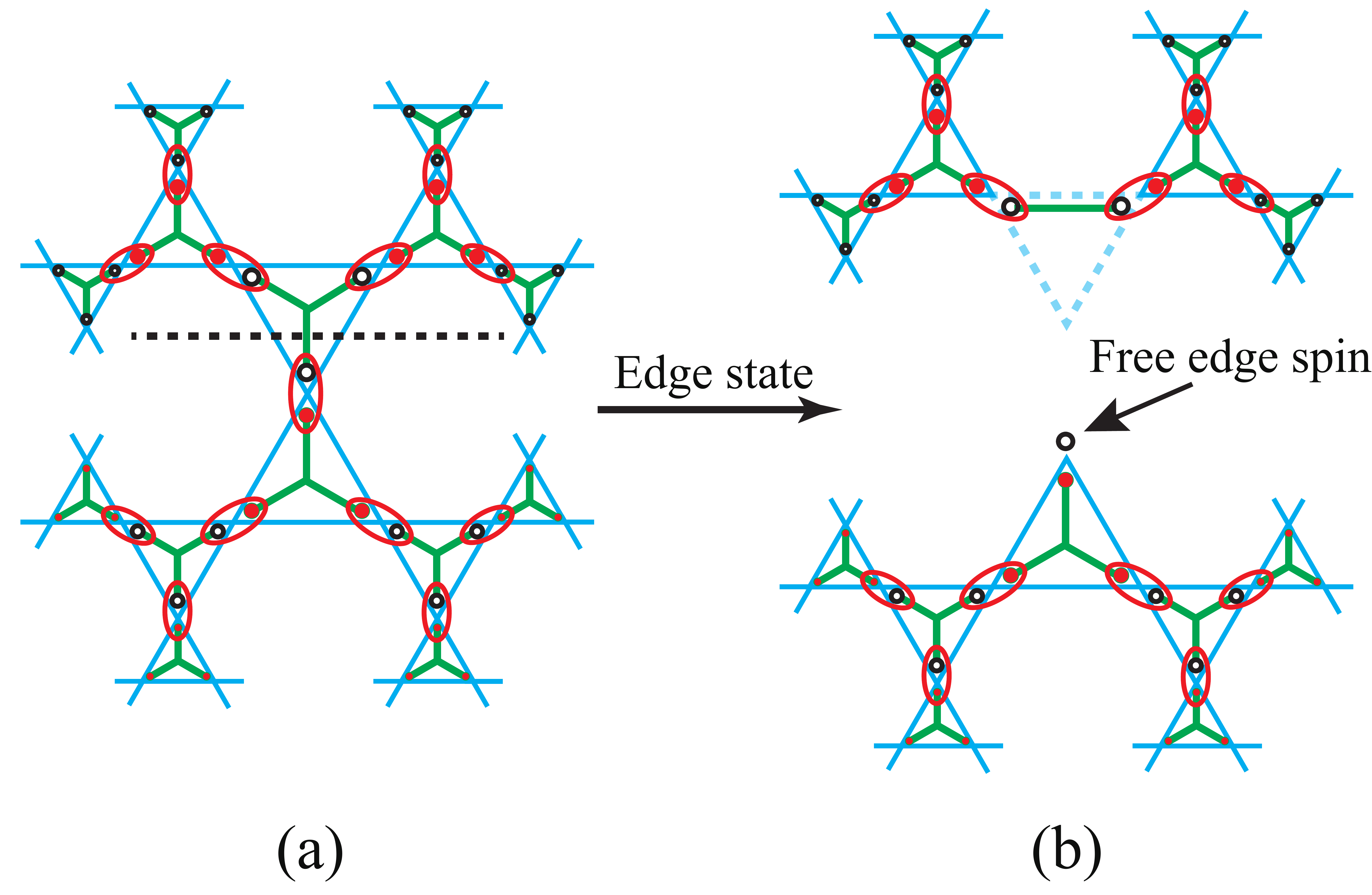,width=8.0cm,angle=0.0}
\end{center}
\caption{(Color online) Schematic representation of (a) a simplex-solid state
and (b) the corresponding edge state. Solid red circles denote virtual spins
$S = n$ and open black circles virtual spins $S = n - 1$ $(S = n)$ for
odd-integer-spin simplex-solid states (even-integer-spin simplex-solid
states). The edge state can be obtained by cutting a single bond of the PESS
and has a free edge spin on its boundary.}
\label{Edge_state}
\end{figure}

As for the AKLT states, the degeneracy of the ES for simplex-solid states
may be understood from the viewpoint of edge states \cite{DP13}. We first
consider the edge states of the odd-integer-spin simplex-solid state,
with $S = 2n - 1$, defined on the Husimi lattice. As shown in
Fig.~\ref{Edge_state}(a), cutting an A-bond of the PESS creates a free
virtual spin $S = n$ at the boundary, represented in Fig.~\ref{Edge_state}(b).
Thus the edge state has a $(2n+1)$-fold degeneracy, which is related directly
to the $(2n+1)$-fold degeneracy of the lowest level of the corresponding ES
(three-fold degeneracy for the $S = 1$ simplex-solid state). Similarly,
cutting a B-bond creates a free virtual spin $S = n - 1$ at the boundary,
which is related to the $(2n-1)$-fold degeneracy of the lowest level of its
corresponding ES (non-degenerate for $S = 1$).

By contrast, for a $S = 2n$ simplex-solid state, the free virtual spin
obtained at the boundary by cutting any bond is $S = n$, giving a
$(2n+1)$-fold degenerate edge state in every case and corresponding to
the $(2n+1)$-fold degeneracy of the lowest levels of the ES on both A- and
B-bonds. This analysis is fully consistent with the above results for $S = 2$
simplex-solid state. We have also verified numerically that the ES of the
exact $S = 3$ and $S = 4$ simplex-solid states confirm this reasoning;
in the $S = 3$ case we find that the lowest levels of the ES for the A-
and B-bonds are respectively five- and three-fold degenerate, while for
$S = 4$ both are five-fold degenerate.

\section{Ground States at Zero Field}

\subsection{$S = 1/2$}
\label{Spin1/2}

We begin our presentation of PESS results for the Heisenberg model on the
Husimi lattice by considering spin quantum number $S = 1/2$. This model was
studied recently by Liu $et$ $al.$~\cite{LRL+14}, who concluded that its
ground state was a featureless quantum spin liquid, with no local
magnetization [Eq.~(\ref{Magnetization})] and no gap, but with exponentially
decaying spin-spin and dimer-dimer correlation functions (these results are
mutually consistent due to the special properties of infinite Bethe-type
lattices, of which the Husimi lattice is an example \cite{LDX12}). However,
we note that the correlation functions and magnetization were calculated by
these authors at finite temperature ($T/J = 0.01$), and thus their results
could be understood simply from the Mermin-Wagner Theorem \cite{MWA66},
which states that there can be no spontaneously broken continuous spin
symmetry at finite temperatures in one and two dimensions.

\begin{figure}[t]
\begin{center}
\epsfig{file=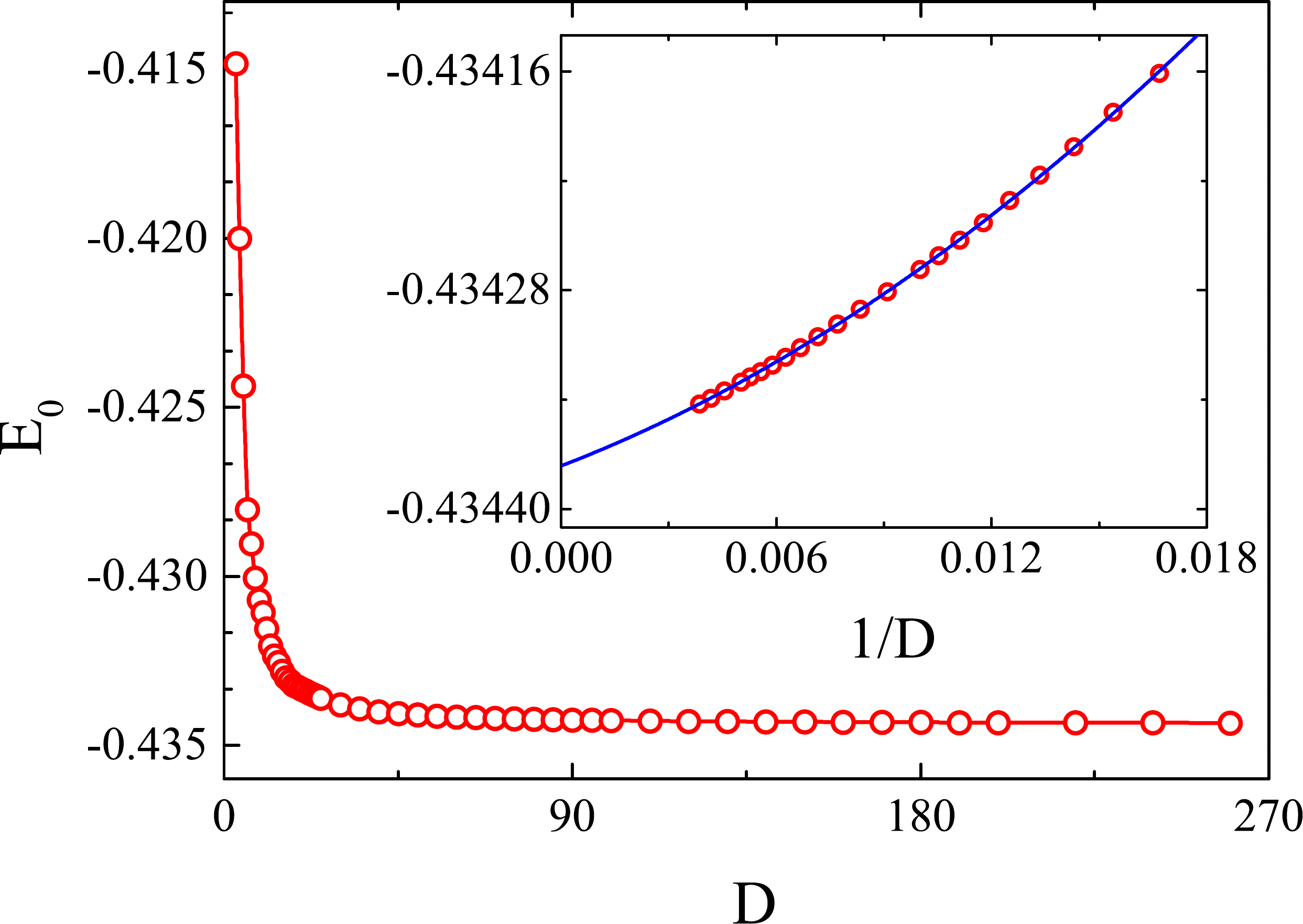,width=8.0cm,angle=0.0}
\end{center}
\caption{(Color online) Energy per site, $E_0$, of the $S = 1/2 $ Heisenberg
model as a function of bond dimension, $D$. The inset shows the energy as a
function of $1/D$, with a polynomial fit shown in blue. The extrapolated
energy is $E^{\infty}_0 = - 0.43438(1)$.}
\label{En_Spin0.5}
\end{figure}

\begin{figure}[t]
\begin{center}
\epsfig{file=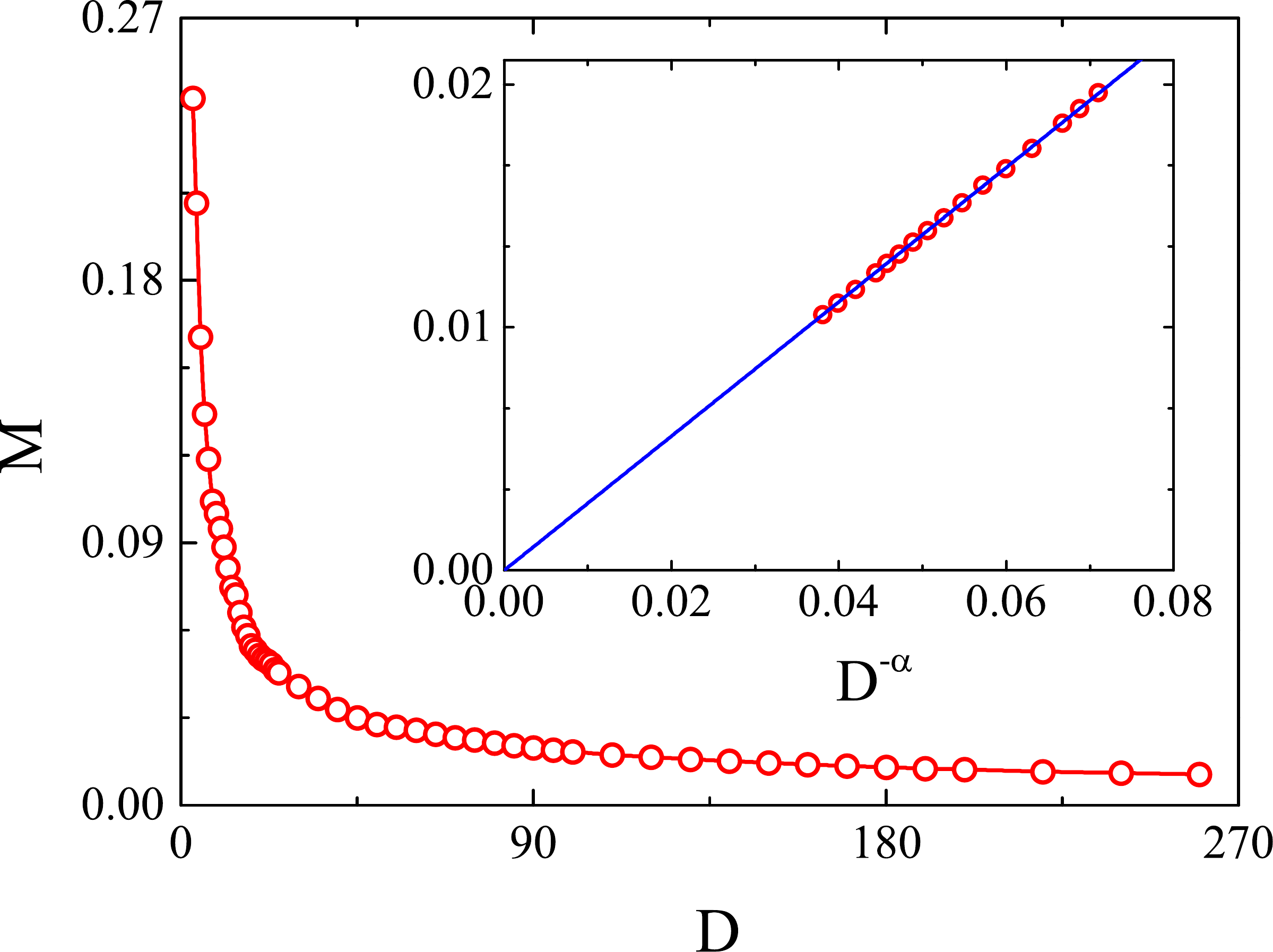,width=8.0cm,angle=0.0}
\end{center}
\caption{(Color online) Local magnetization, $M$, of the $S = 1/2$ Heisenberg
model as a function of bond dimension, $D$. The inset shows $M$ as a function
of $D^{-\alpha}$, with $\alpha = 0.588(2)$. The intercept of the linear fit
(solid blue line) is $M (\infty) = 0.00000(4)$.}
\label{Ms_Spin0.5}
\end{figure}

Here we investigate the properties of the model at zero temperature and
perform careful extrapolations to the limit of infinite bond dimension, $D$.
The ground-state energy per site, $E_0(D)$, is shown in Fig.~\ref{En_Spin0.5}
for values up to $D = 260$; its extrapolation to infinite $D$ (inset,
Fig.~\ref{En_Spin0.5}), obtained by a quadratic fit, is $E^{\infty}_0 =
 - 0.43438(1)$, and thus agrees within the error bars with the result
of Ref.~\cite{LRL+14}. If we consider the magnetization order parameter,
$M$ (\ref{Magnetization}), shown in Fig.~\ref{Ms_Spin0.5}, we find that
this is finite at every value of $D$, but decreases as $D$ increases.
We also confirm that the spin orientations conform to the expected
$120$-degree N\'eel order (Fig.~{\ref{AF_120}}) and that the spin
magnitudes, $\sqrt{\langle S_{i}^{x} \rangle^{2} + \langle S_{i}^{y} \rangle^{2}
 + \langle S_{i}^{z} \rangle^{2}}$, are identical at every site within
numerical error.

The key question is how this magnetization behaves in the limit of infinite
$D$. On logarithmic axes, our $M(D)$ data from $D = 90$ to 260 fall on a
perfectly straight line, whose gradient we obtain as $\alpha = - 0.588(2)$.
The inset of Fig.~\ref{Ms_Spin0.5} shows this purely algebraic functional
form, $M \propto D^{-\alpha}$. We do not show data for spin-spin or dimer-dimer
correlation functions because, as noted above, they contain no new information
about the gapped or gapless nature of the system. The key issue is instead
the question of whether the gapless, algebraic system may in fact have
long-range magnetic order. Extrapolation of our results to the limit $D
\rightarrow \infty$ yields the result $M(\infty) = 0.00000(4)$. To summarize,
we have taken extreme care to obtain the most precise, high-$D$ data at zero
temperature and to extrapolate it following the most accurate possible
protocols, which leads us to conclude that the ordered moment vanishes, and
does so algebraically. Taking this result in combination with the polynomial
convergence of the ground-state energy (Fig.~\ref{En_Spin0.5}), we therefore
draw one of the most important conclusions of the present study, that the
true ground state of the $S = 1/2$ Heisenberg model on the triangular Husimi
lattice is a gapless, non-magnetic state, i.e.~an algebraic spin liquid.

\begin{figure}[t]
\begin{center}
\epsfig{file=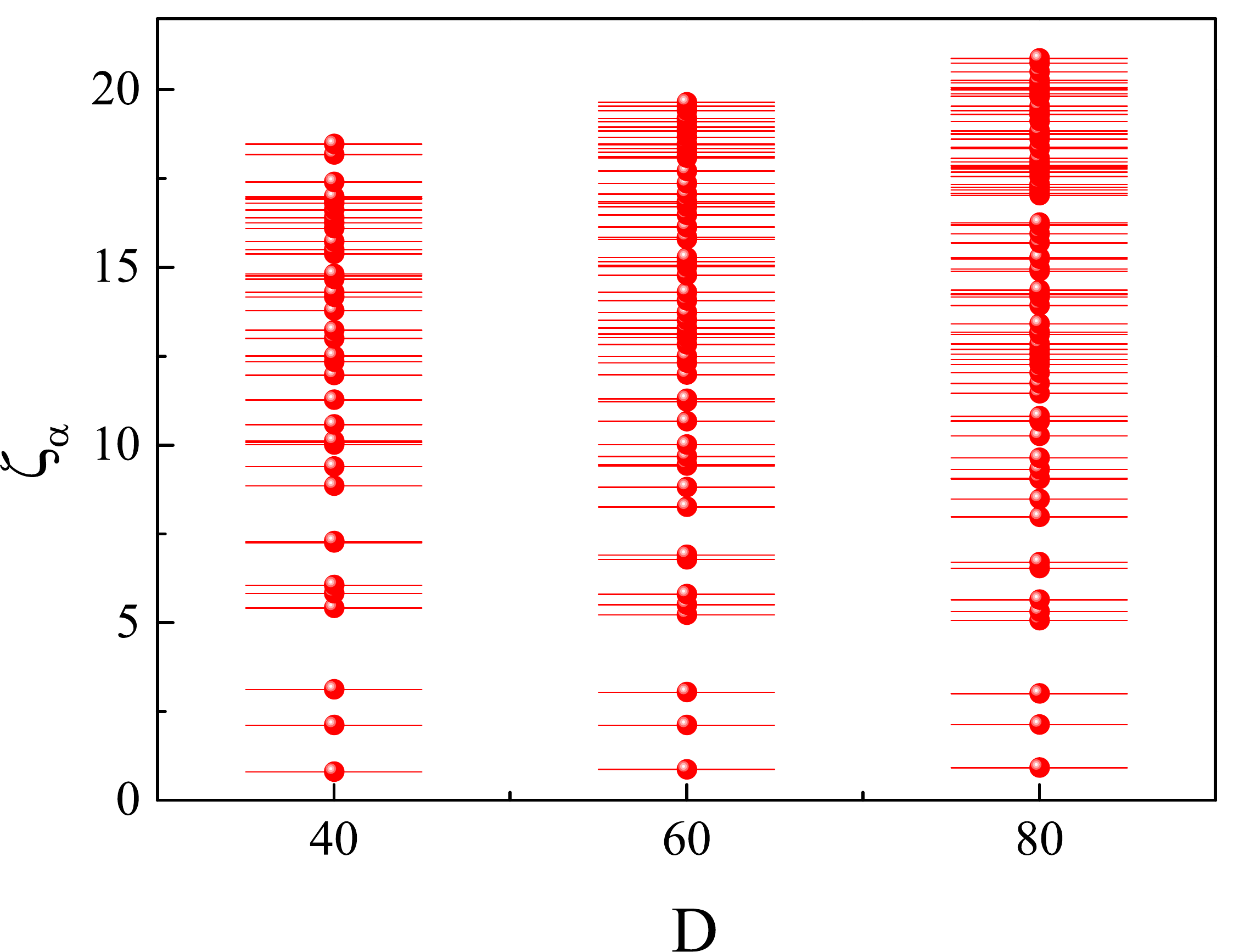,width=8.0cm,angle=0.0}
\end{center}
\caption{(Color online) Entanglement spectra of the $S = 1/2 $ Heisenberg
model with $D = 40$, 60, and 80. The number of dots on each level denotes
its degeneracy; every low-lying level in the spectrum is non-degenerate.}
\label{ES_Spin0.5}
\end{figure}

For completeness we show in Fig.~\ref{ES_Spin0.5} the ES of the $S = 1/2$
Heisenberg model on the Husimi lattice. Clearly all levels of the ES are
non-degenerate, with the lowest three well separated from the others. We
return below to an interpretation of these results in the light of our
further findings.

\subsection{$S = 1$}
\label{Spin1}

\begin{figure}[t]
\begin{center}
\epsfig{file=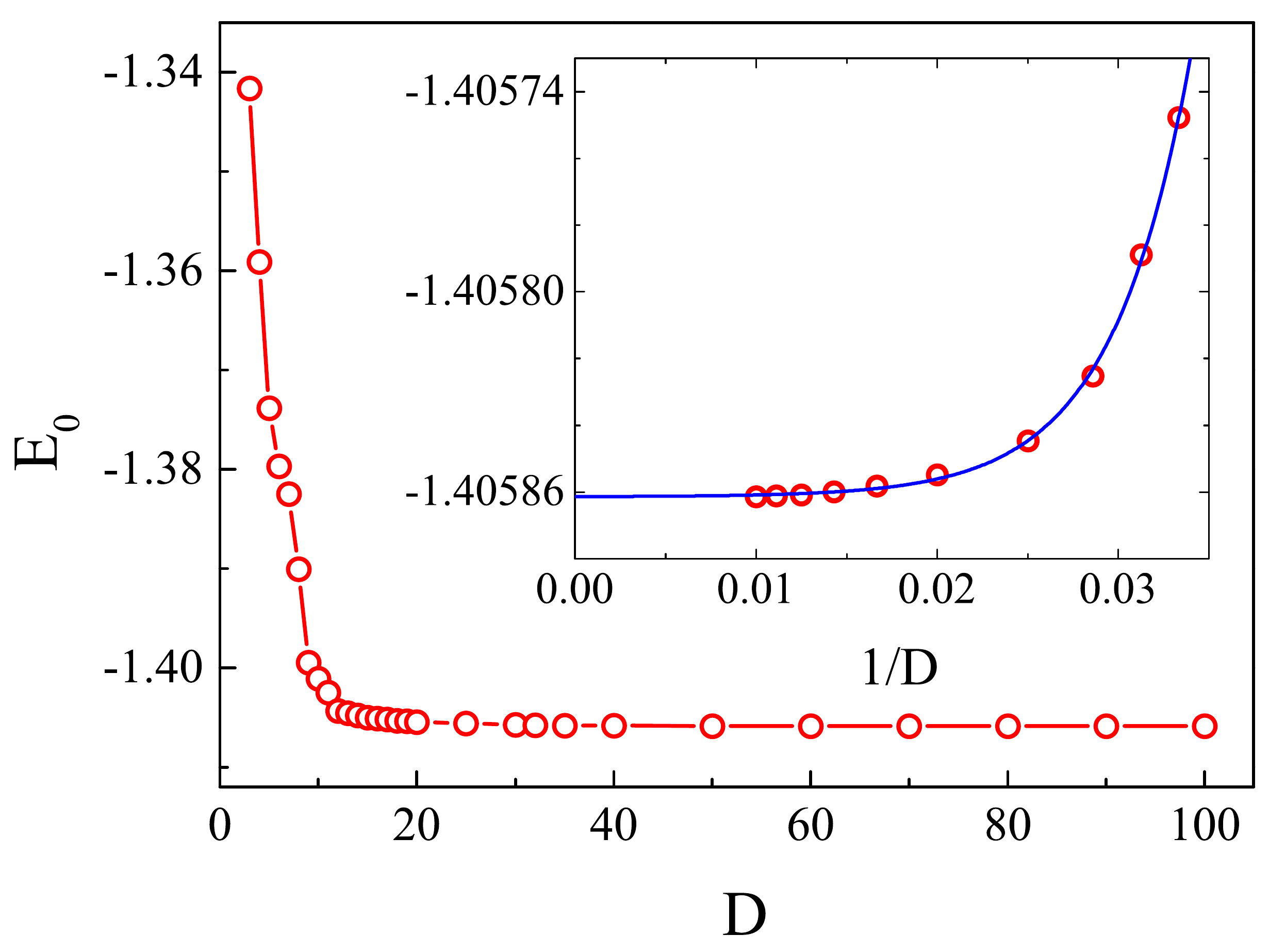,width=8.0cm,angle=0.0}
\end{center}
\caption{(Color online) Energy per site, $E_0$, of the $S = 1 $ Heisenberg
model as a function of bond dimension, $D$. The inset shows the energy as
a function of $1/D$, with an exponential fit shown in blue. The extrapolated
energy is $E^{\infty}_0 = - 1.40586(1)$.}
\label{En_Spin1.0}
\end{figure}

\begin{figure}[t]
\begin{center}
\epsfig{file=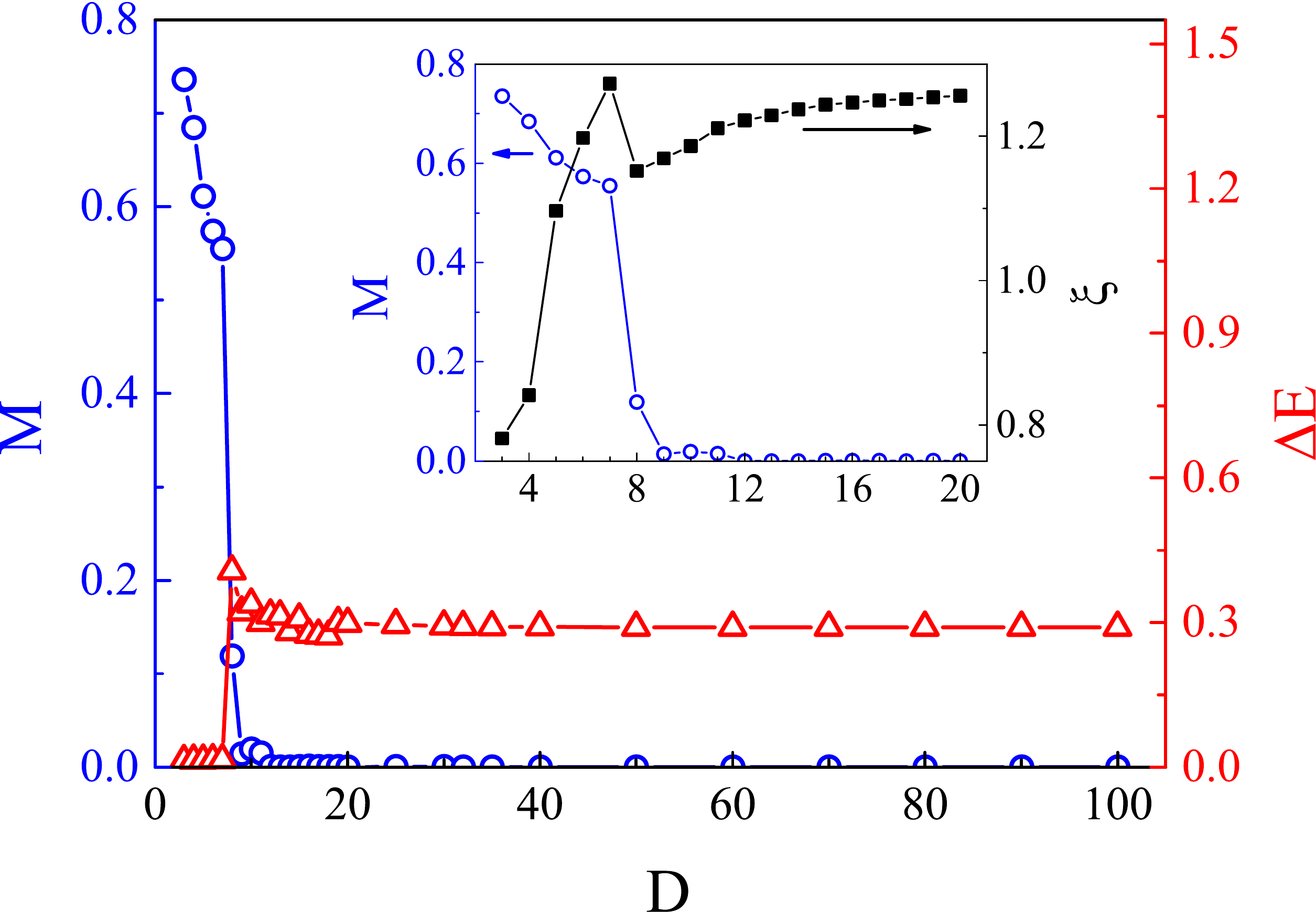,width=8.5cm,angle=0.0}
\end{center}
\caption{(Color online) Trimerization parameter $\Delta E = 2|E_{\triangle}
- E_{\nabla}|/3$ (open red triangles), where $E_{\triangle(\nabla)}$ is the average
energy of an up-triangle (down-triangle), and local magnetization, $M$ (open
blue circles), both shown as a function of bond dimension, $D$. Finite
trimerization and vanishing magnetization occur simultaneously when $D \ge 8$.
The correlation length, $\xi$ (solid black squares), also drops abruptly at
$D = 8$, as shown in the inset, further confirming the occurrence of a phase
transition at the bond dimension $D_c = 8$. The trimerization converges to a
constant value, $\Delta E = 0.29021(1)$, at large $D$.}
\label{Ms_Spin1.0}
\end{figure}

Turning to the $S = 1$ Heisenberg model on the Husimi lattice, in
Fig.~\ref{En_Spin1.0} we present our results for the ground-state energy
per site, $E_0$, up to $D = 100$. Once again $E_0$ decreases monotonically
with $D$, converging rapidly (inset, Fig.~\ref{En_Spin1.0}) to $E_0 = -
1.405861(1)$ in the infinite-$D$ limit. However, significantly more insight
into the nature of the ground state may be obtained from the energy difference
between between up- and down-triangles, $\Delta E = 2| E_{\triangle} -
E_{\nabla}|/3$, which we show as a function of $D$ in Fig.~\ref{Ms_Spin1.0}.
$\Delta E$ may be considered as a type of trimerization ``order parameter,''
and undergoes a rapid onset at $D = 8$. The magnetic order parameter, $M$,
also shown in Fig.~\ref{Ms_Spin1.0}, undergoes an equally rapid fall to zero
at the same value, while the correlation length (inset, Fig.~\ref{Ms_Spin1.0})
also drops abruptly. These features all demonstrate that a phase transition
from a magnetically ordered state to a nonmagnetic, trimerized state occurs
at the bond dimension $D_c = 8$, which we note corresponds to the minimum
$D$ required to describe a state of finite trimerization.

\begin{figure}[t]
\begin{center}
\epsfig{file=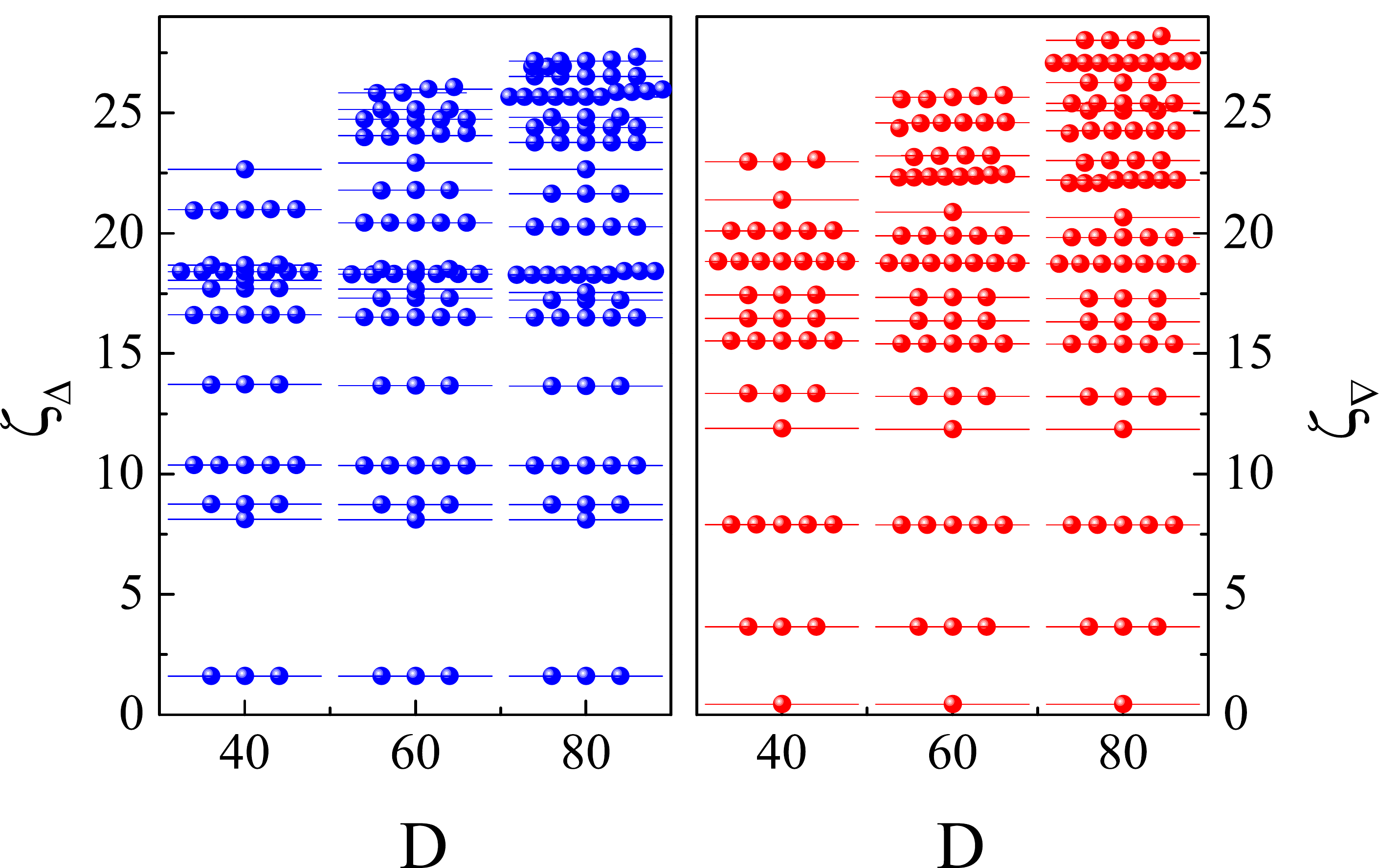,width=8.0cm,angle=0.0}
\end{center}
\caption{(Color online) Entanglement spectra of the $S = 1$ Heisenberg
model with $D = 40$, 60, and 80 on A-bonds (left) and B-bonds (right). The
number of dots on each level denotes its degeneracy. Three-fold degeneracy
of the lowest A-bond levels indicates simplex singlet entanglement within the
up-triangles, whereas the non-degenerate B-bond levels indicate its absence
within the down-triangles.}
\label{ES_Spin1.0}
\end{figure}

The energy difference, $\Delta E$, converges to a constant value [$0.29021(1)$]
as $D$ becomes large, indicating that the trimerized state persists as the
true ground state. To confirm this result, we calculated the ES of the $S = 1$
model, which is shown in Fig.~\ref{ES_Spin1.0}. Clearly the lowest-lying
level on the A-bond is three-fold degenerate while that on the B-bond is
non-degenerate. These are exactly the properties of the non-uniform simplex
solid state of the $S = 1$ model discussed in Subsec.~\ref{Simplex_ES}. Thus
the ES verifies that the ground state of the $S = 1$ Heisenberg model on the
Husimi lattice is a trimerized simplex-solid state with a spontaneous
breaking of lattice inversion symmetry. The most direct demonstration that
this state has a finite gap is obtained from the longitudinal magnetization
(Sec.~IID3), which we discuss in detail in Sec.~IV.

\begin{figure}[t]
\begin{center}
\epsfig{file=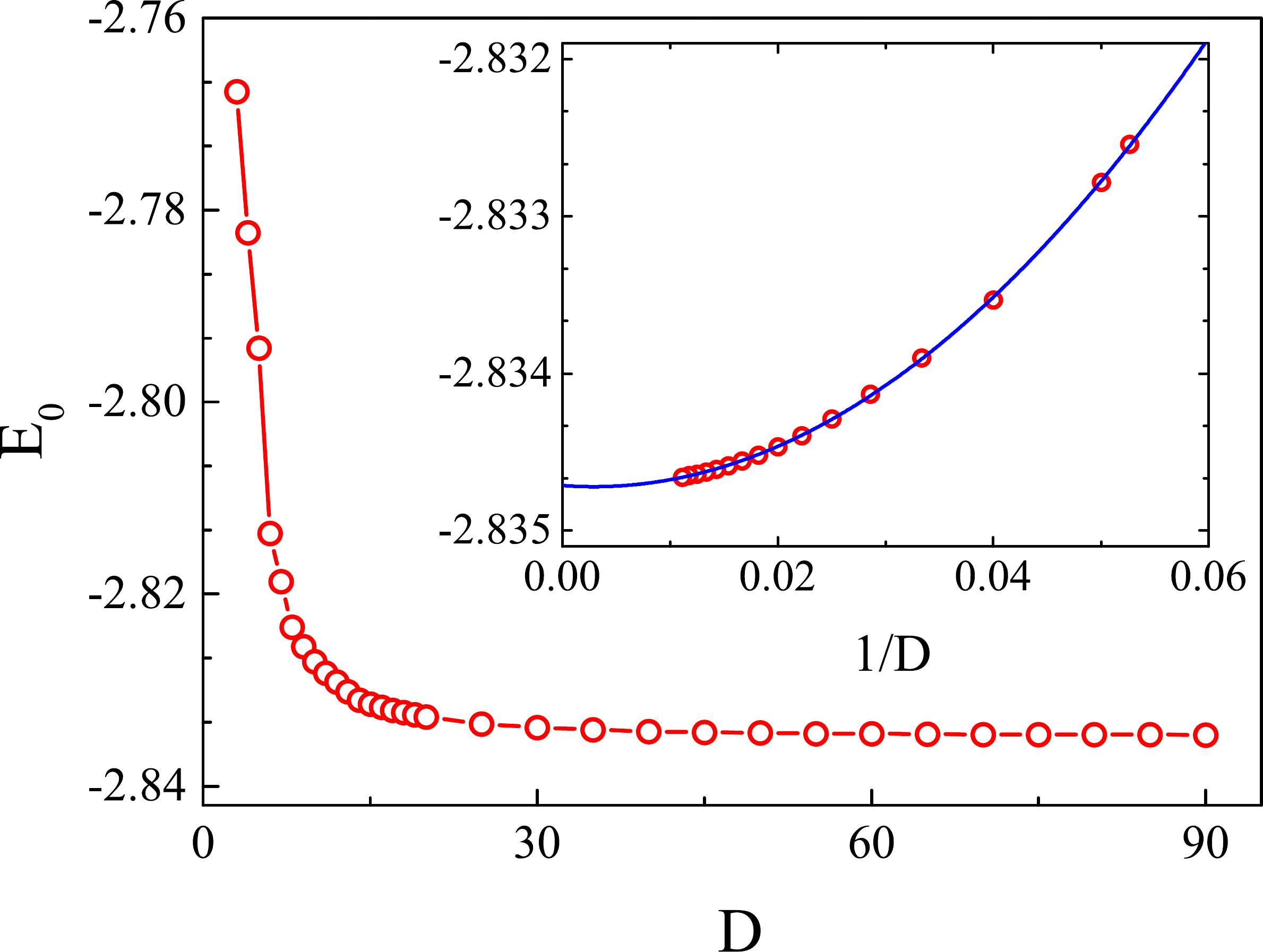,width=8.0cm,angle=0.0}
\end{center}
\caption{(Color online) Energy per site, $E_0$, of the $S = 3/2 $ Heisenberg
model as a function of bond dimension, $D$. The inset shows the energy as a
function of $1/D$, with a polynomial fit shown in blue. The extrapolated energy
is $E^{\infty}_0 = - 2.83471(1)$.}
\label{En_Spin1.5}
\end{figure}

As noted in Sec.~I, the $S = 1$ Heisenberg model on the kagome lattice has
recently attracted strong attention. Older proposals for the ground state,
including the hexagonal singlet solid \cite{Hida00} and the resonating AKLT
loop state \cite{YFQ10,LYC+14}, appear to have been supplanted by a trimerized
simplex-solid state \cite{CCW09,LLW+15,CL15}, which has the best variational
energy, $E_0^k = - 1.4116(4)$ \cite{LLW+15}. This is a symmetry-broken state
with trimerization order, which as above can be defined by the difference of
the average energies between up- and down-triangle simplices, quoted in
Ref.~\cite{LLW+15} as 0.261, or approximately 19\% of $E_0^k$. Here we have
shown that the same model on the Husimi lattice has exactly the same type of
ground state, a trimerized simplex solid, with a remarkably similar energy,
$E_0 = - 1.40586$, and a trimerization parameter of approximately 20.6\%. Thus
one may conclude that the vast majority of energetic effects on the two
lattices are dominated by extremely local processes.

\subsection{$S = 3/2$}
\label{Spin3/2}

\begin{figure}[t]
\begin{center}
\epsfig{file=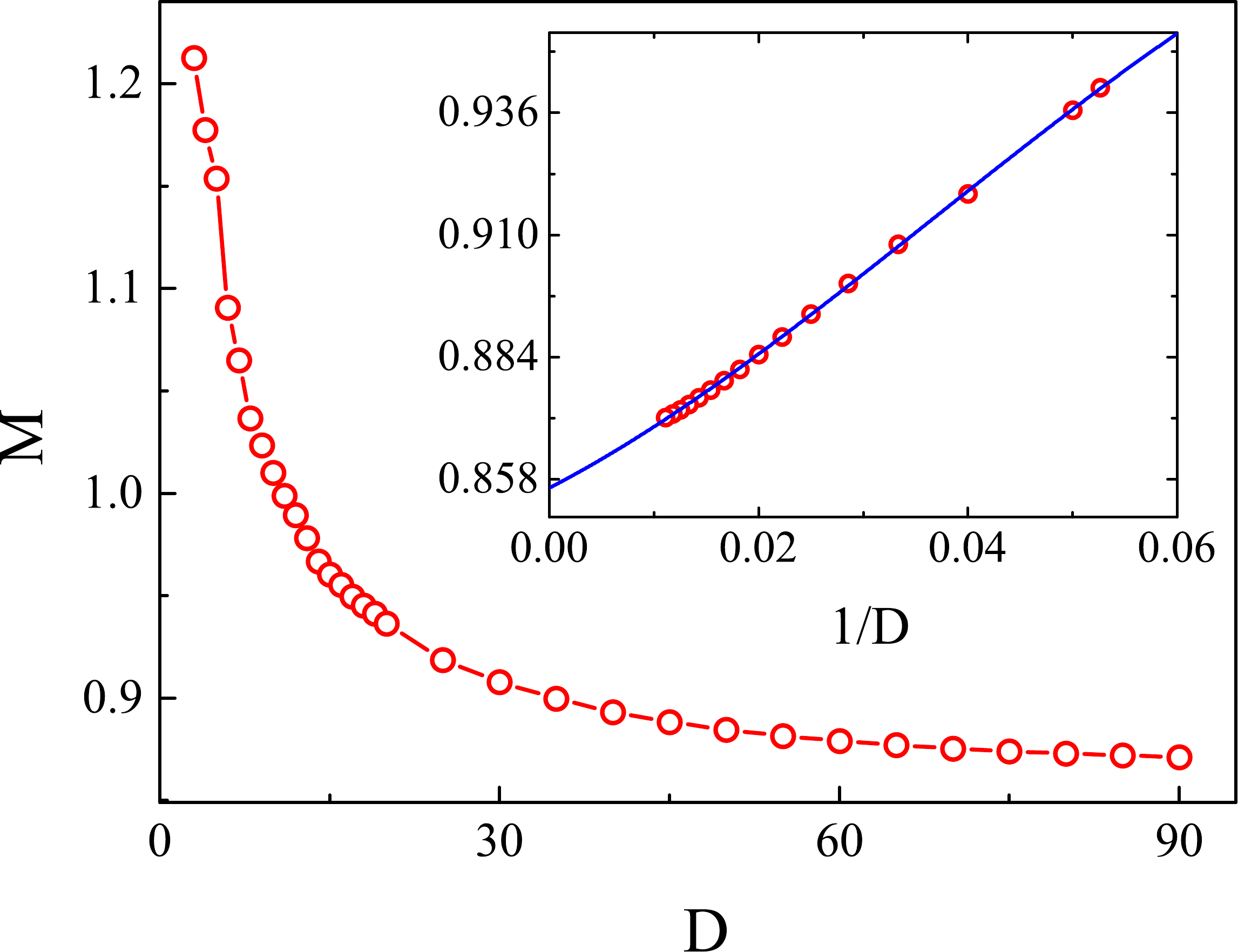,width=8.0cm,angle=0.0}
\end{center}
\caption{(Color online) Local magnetization, $M$, of the $S = 3/2$ Heisenberg
model as a function of bond dimension, $D$. The inset shows $M$ as a function
of $1/D$, with a polynomial fit shown in blue. The extrapolated $M$ has a
finite value at infinite $D$, $M(\infty) = 0.856(3)$.}
\label{Ms_Spin1.5}
\end{figure}

For the $S = 3/2$ case, we find as for $S = 1/2$ that the ground-state energy
per site converges algebraically [to $E^{\infty}_0 = - 2.83471(1)$] with
increasing bond dimension, as shown in Fig.~\ref{En_Spin1.5}; such convergence
behavior indicates that the system is gapless. However, the magnetic order
parameter, shown in Fig.~\ref{Ms_Spin1.5}, converges not to zero at large $D$
but to a finite value, $M(\infty) = 0.856(3)$. This robust magnetization is
approximately $57\%$ of the classical value for an $S = 3/2$ system. N\'eel
order is of course consistent with gapless excitations and algebraic
convergence of $E(D)$, but demonstrates clearly that the physics of the
$S = 3/2$ Husimi lattice has more in common with high-dimensional
antiferromagnets than with the physics of spin chains and the Haldane
conjecture. We defer a discussion of the ES for this case to Sec.~IIIE.

\begin{figure}[t]
\begin{center}
\epsfig{file=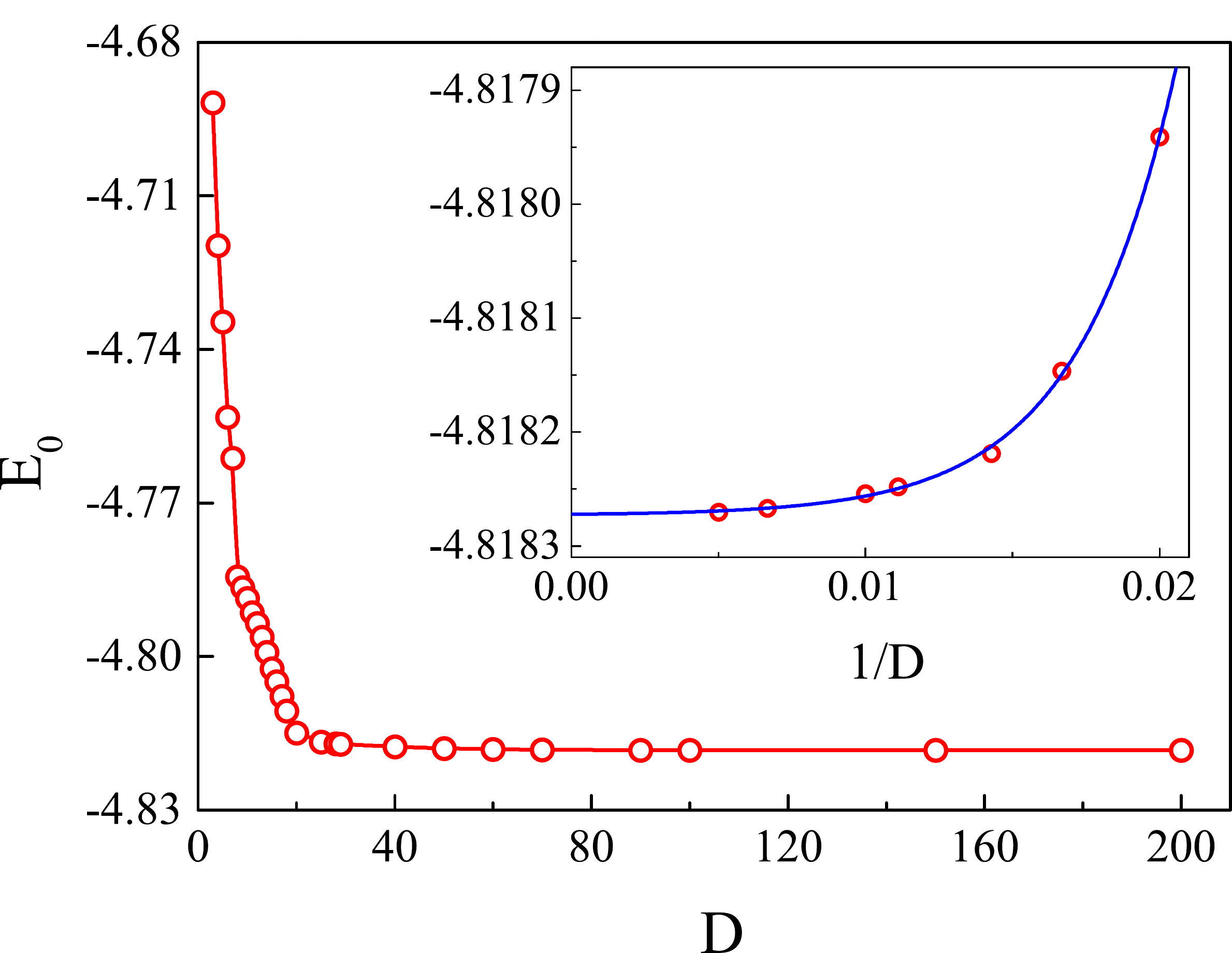,width=8.0cm,angle=0.0}
\end{center}
\caption{(Color online) Energy per site, $E_0$, of the $S = 2 $ Heisenberg
model as a function of bond dimension, $D$. The inset shows the energy as a
function of $1/D$, with as exponential fit shown in blue. The extrapolated
energy is $E^{\infty}_0 = - 4.8185(4)$.}
\label{En_Spin2.0}
\end{figure}

\subsection{$S = 2$}
\label{Spin2}

\begin{figure}[t]
\begin{center}
\epsfig{file=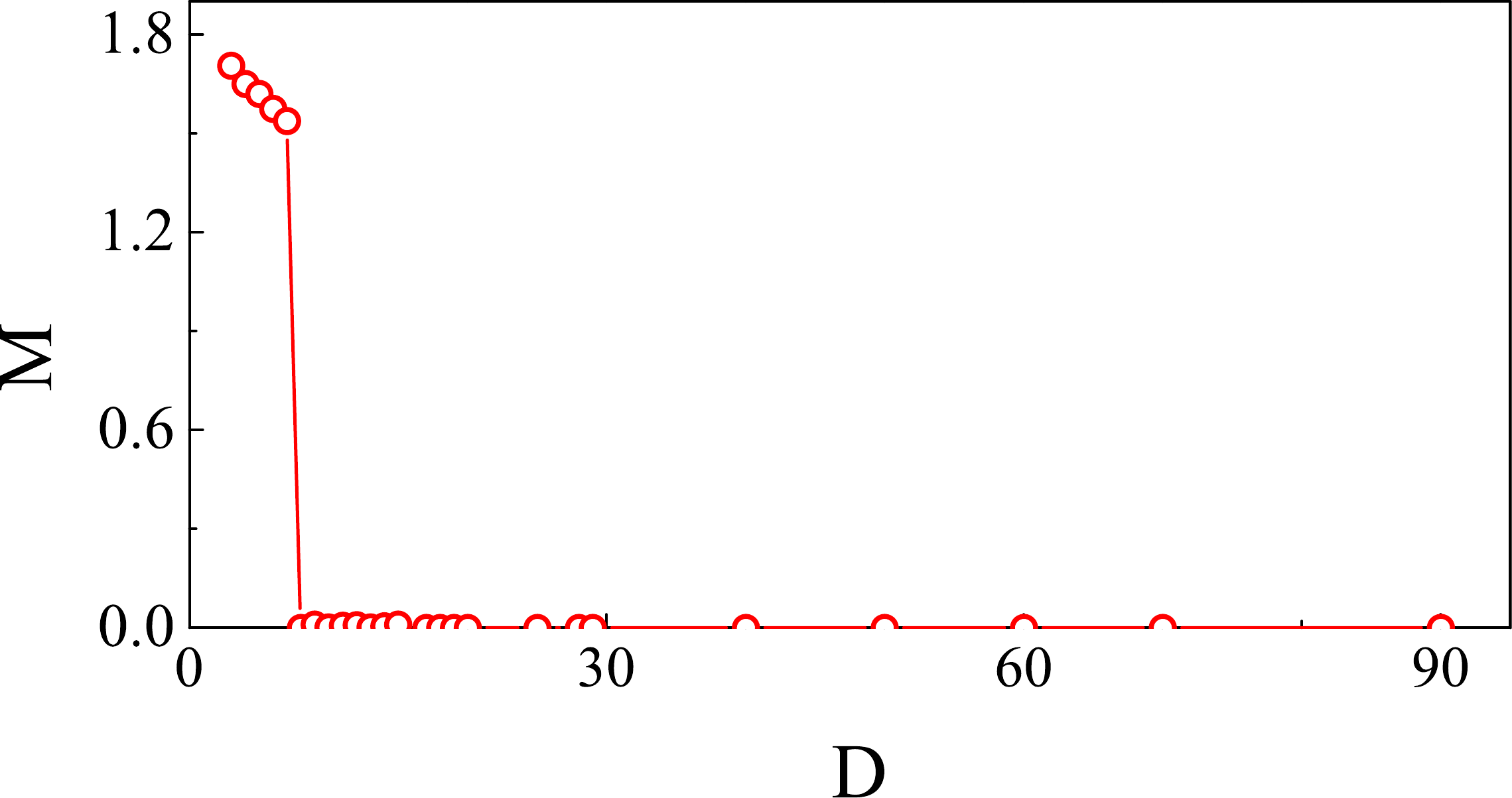,width=8.0cm,angle=0.0}
\end{center}
\caption{(Color online) Local magnetization, $M$, of the $S = 2 $ Heisenberg
model as a function of bond dimension, $D$. Sudden vanishing of $M$ for bond
dimension $D \ge 8$ indicates a transition to a non-magnetic state.}
\label{Ms_Spin2.0}
\end{figure}

Turning to the Heisenberg model with $S = 2$, the ground-state energy per
site (Fig.~\ref{En_Spin2.0}) again converges monotonically and rapidly to
$E_0^{\infty} = - 4.8185(4)$; as for $S = 1$, we fit only the higher-$D$ values
of this somewhat stepwise convergence to an exponential form. Also as for $S
 = 1$, we find again that the magnetization, $M$, vanishes suddenly for $D \ge
8$, as shown in Fig.~\ref{Ms_Spin2.0}, proving that the ground state in this
case is non-magnetic. For a direct calculation of the spin gap, the
longitudinal magnetization (Sec.~IID3) is shown in Sec.~IV.

\begin{figure}[t]
\begin{center}
\epsfig{file=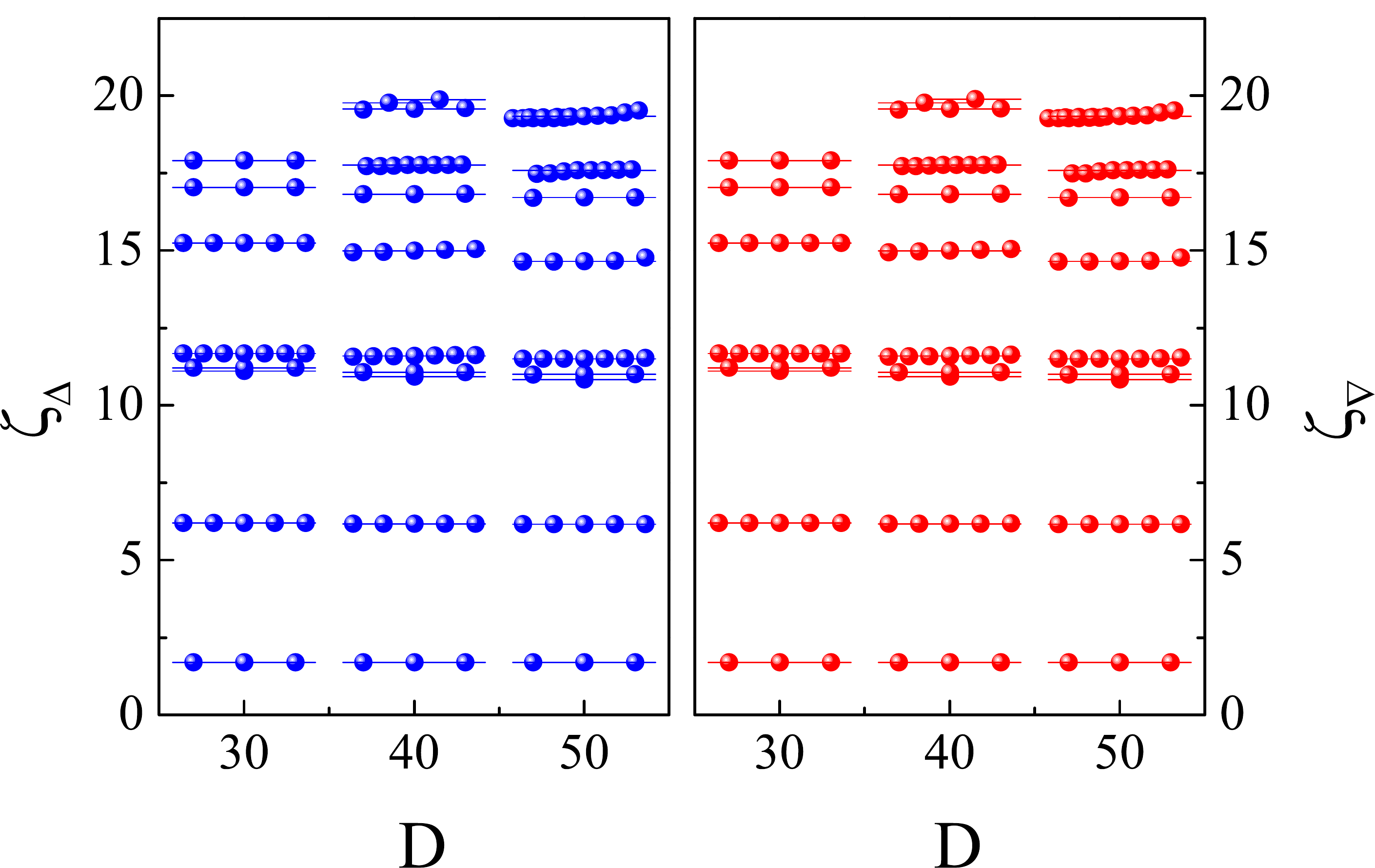,width=8.0cm,angle=0.0}
\end{center}
\caption{(Color online) Entanglement spectra of the $S = 2$ Heisenberg model
with $D = 30$, 40, and 50 on A-bonds (left) and B-bonds (right). The number
of dots on each level denotes its degeneracy. Three-fold degeneracy of the
lowest-lying levels for every bond in the system indicates singlet
entanglement within every simplex (triangle).}
\label{ES_Spin2.0}
\end{figure}

Further insight into the nature of this spin liquid is obtained from the
ES, shown in Fig.~\ref{ES_Spin2.0}. Unlike the $S = 1$ case, here the lowest
levels on both A- and B-bonds are three-fold degenerate, demonstrating the
presence of a simplex singlet on every simplex in the system. These results
are fully consistent with those obtained for the exact $S = 2$ simplex-solid
state discussed in Subsec.~\ref{Simplex_ES}, and thus the ES indicates that
the ground state for $S = 2$ is a uniform simplex-solid state.

\begin{table}[b]
\caption{Expected values of spin projection operators at each simplex for the
$S = 2$ Heisenberg model on the Husimi lattice, calculated with bond dimension
$D = 40$.}
\begin{tabular}{|c|c|c|}\hline
 & $\langle \Psi |P_{J}^{\triangle}| \Psi \rangle$    & $\langle \Psi
|P_{J}^{\nabla}| \Psi \rangle$    \\ \hline
$J = 0$  & 0.1189(1)    & 0.1189(1)    \\ \hline
$J = 1$  & 0.5391(1)    & 0.5391(1)    \\ \hline
$J = 2$  & 0.2803(1)    & 0.2803(1)    \\ \hline
$J = 3$  & 0.0564(1)    & 0.0564(1)    \\ \hline
$J = 4$  & 0.00518(2)   & 0.00518(2)   \\ \hline
$J = 5$  & 0.000216(2)  & 0.000216(2)  \\ \hline
$J = 6$  & 0.000004(1)  & 0.000004(1)  \\ \hline
\hline
\end{tabular}
\label{table1}
\end{table}

For further confirmation of the properties of this simplex solid, we
calculate the expectation value of the spin projection operator, $P_{J}^{\alpha}$
[Eq.~(\ref{SPO})], which projects the state at each simplex $\alpha$ onto
a state of total spin $J$. From the values given in Table~\ref{table1},
it is clear that the quantity $\langle \Psi |P_{0}^{\alpha} + P_{1}^{\alpha}
 + P_{2}^{\alpha} + P_{3}^{\alpha}| \Psi \rangle = 0.9947(1)$ is very close to
unity, i.e.~it is very improbable that the total spin at each simplex could
exceed $3$. Following the analysis of Sec.~\ref{SimplexSS}, there is therefore
a simplex singlet [which belongs to the $S = 0$ subspace of $(1 \otimes 1
\otimes 1)$] on every simplex with near-unit probability. Thus the ground
state of the $S = 2$ Heisenberg model on the Husimi lattice lies very close
to the exact simplex-solid state.

\subsection{Higher spin: $S = 5/2$, 3, $7/2$, and 4}
\label{Higher_S}

From the results of the previous four subsections, the spin-$S$ Heisenberg
model on the Husimi lattice provides one example of a gapless spin liquid,
one uniform simplex-solid state, one non-uniform simplex solid, and one
ordered antiferromagnet. These results imply very strong variability and
an equally strong ``odd-even'' effect between integer and half-odd-integer
spins. The complete lack of systematics to date mandates continuing the
study to higher $S$ values.

\begin{figure}[t]
\begin{center}
\epsfig{file=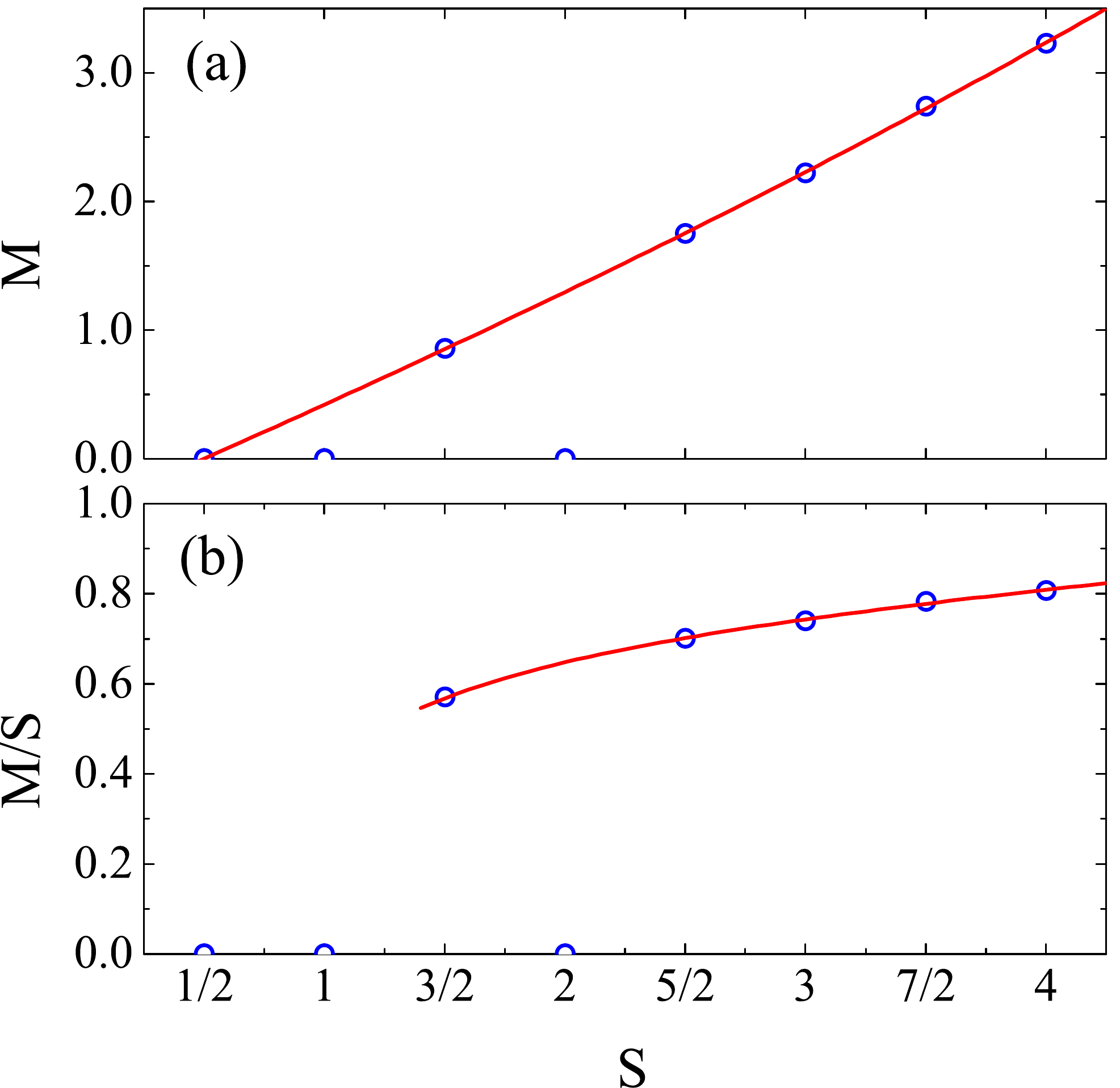,width=8.0cm,angle=0.0}
\end{center}
\caption{(Color online) (a) Extrapolated magnetization $M (\infty)$
as a function of spin quantum number $S$ for $S \le 4$. (b) Ratio
of extrapolated magnetization $M$ to classical spin magnitude $S$.
Red lines are guides to the eye.}
\label{Ms_tot}
\end{figure}

\begin{figure}[t]
\begin{center}
\epsfig{file=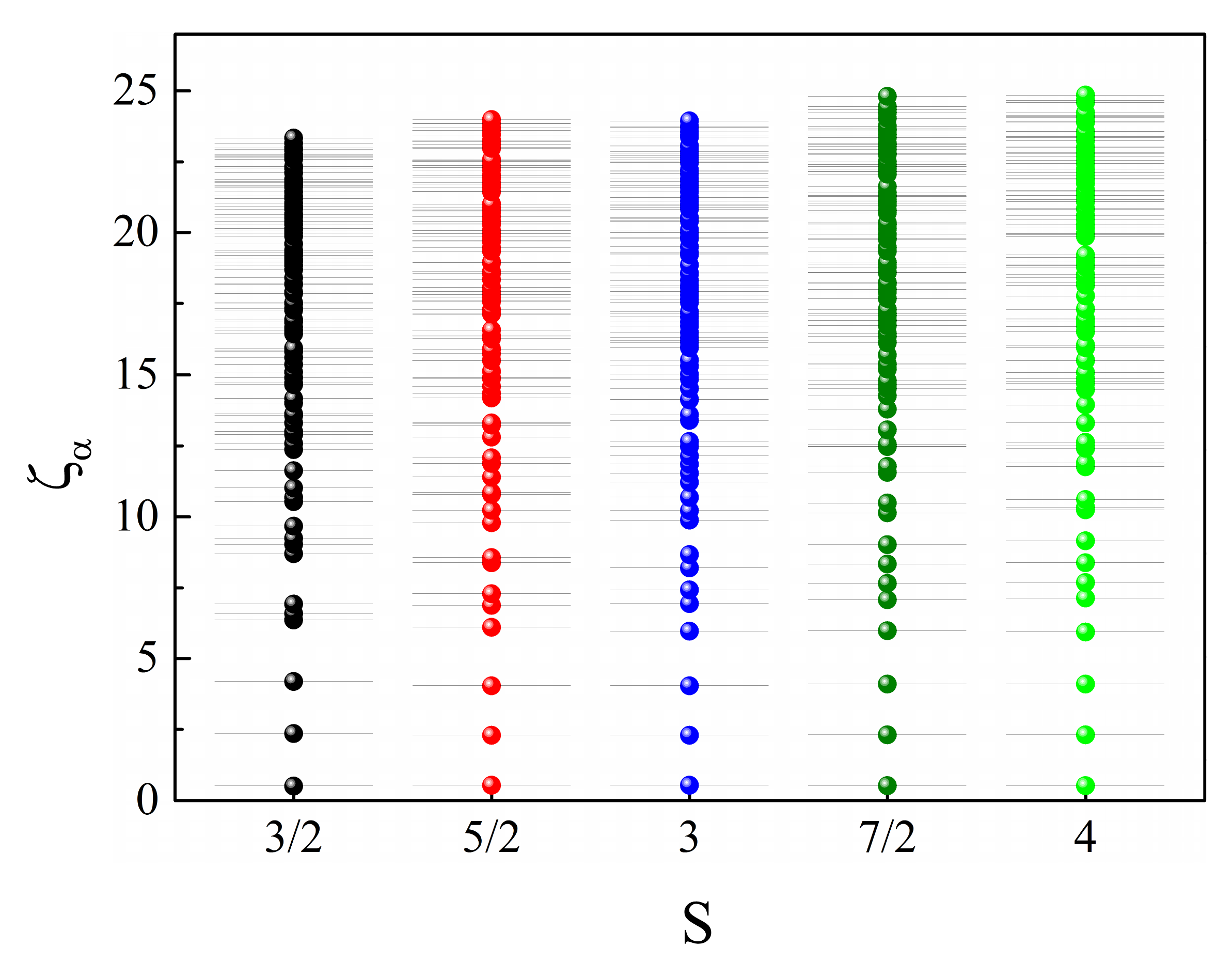,width=8.0cm,angle=0.0}
\end{center}
\caption{(Color online) Entanglement spectra of the Heisenberg model on
the Husimi lattice for $S = 3/2$, 5/2, 3, 7/2, and $4$, calculated with
bond dimension $D = 90$. All bonds are equivalent and all low-lying levels
in the spectra are non-degenerate.}
\label{ES_tot}
\end{figure}

However, by considering the next four $S$ values up to $S = 4$, we find that
the ground states are all antiferromagnetically ordered with the classical
$120$-degree N\'eel configuration. The spontaneous magnetization values $M$
are shown as a function of $S$ in Fig.~\ref{Ms_tot}. Beyond $S = 2$, $M$
clearly tends monotonically towards its classical value, $S$, with no evidence
even for alternation effects related to the integer or half-odd-integer nature
of the quantum spin. Thus ``quantum effects,'' meaning the relevance of
quantum mechanical fluctuations, really are limited to small $S$ values
($S = 1/2$, $1$, and $2$), before classical physics becomes dominant.
This is equally true for the half-odd-integer series, where only $S = 1/2$
is ``quantum enough'' to remain disordered while $S = 3/2$ is well
antiferromagnetically ordered, as it is for the integer series, where
only $S = 1$ and 2 have simplex-solid states more favorable than classical
order.

\begin{figure}[t]
\begin{center}
\epsfig{file=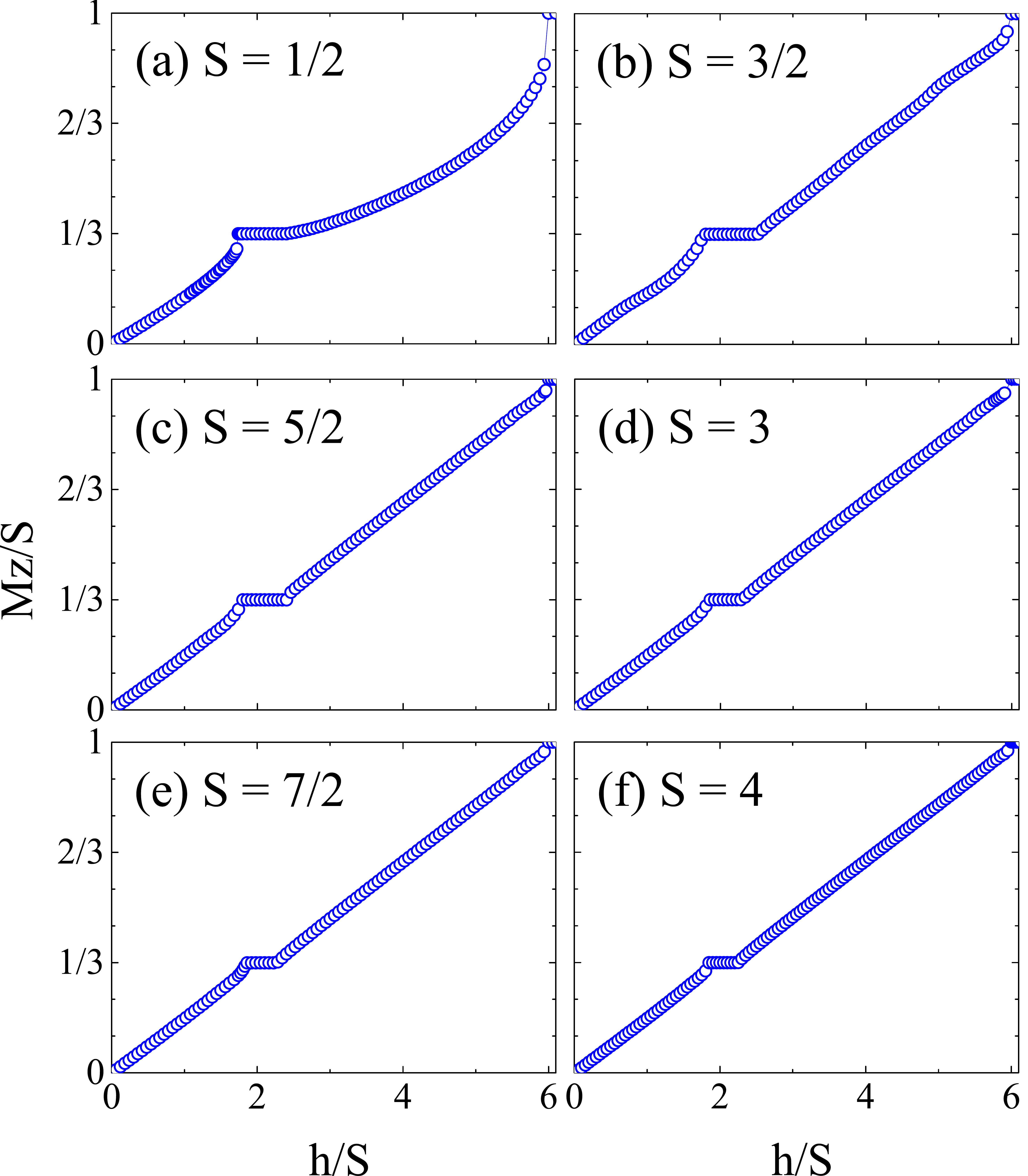,width=8.5cm,angle=0.0}
\end{center}
\caption{(Color online) Longitudinal magnetization, $M_z$, normalized by
its saturation value, $S$, as a function of external magnetic field, $h$,
for (a) $S = 1/2$, (b) $S = 3/2$, (c) $S = 5/2$, (d) $S = 3$, (e) $S =
7/2$, and (f) $S = 4$; calculations performed with bond dimension $D = 30$.
In every case, $M_z$ rises linearly from zero with applied field, and a
magnetization plateau is present at $1/3$ of the saturation value.}
\label{Mh_tot}
\end{figure}

Information about the concomitant entanglement can be obtained from
the ES, which is shown in Fig.~\ref{ES_tot} for all spins $3/2 \le S
\le 4$. In contrast to the simplex-solid states found for $S = 1$ and 2
(Figs.~\ref{ES_Spin1.0} and \ref{ES_Spin2.0}), the structure of the ES is
very simple, with all low-lying levels being non-degenerate. This indicates
that the antiferromagnetically ordered state has no many-body entanglement
and is effectively just a product state with short-range entanglement only.
This property is a common characteristic for all magnetically ordered
phases and allows us also to interpret the results for the $S = 1/2$ case
(Fig.~\ref{ES_Spin0.5}), when we recall that this state has finite N\'eel
order for all finite values of $D$.

\section{Ground States with Applied Magnetic Field}

\subsection{Longitudinal Magnetization}

\begin{figure}[t]
\begin{center}
\epsfig{file=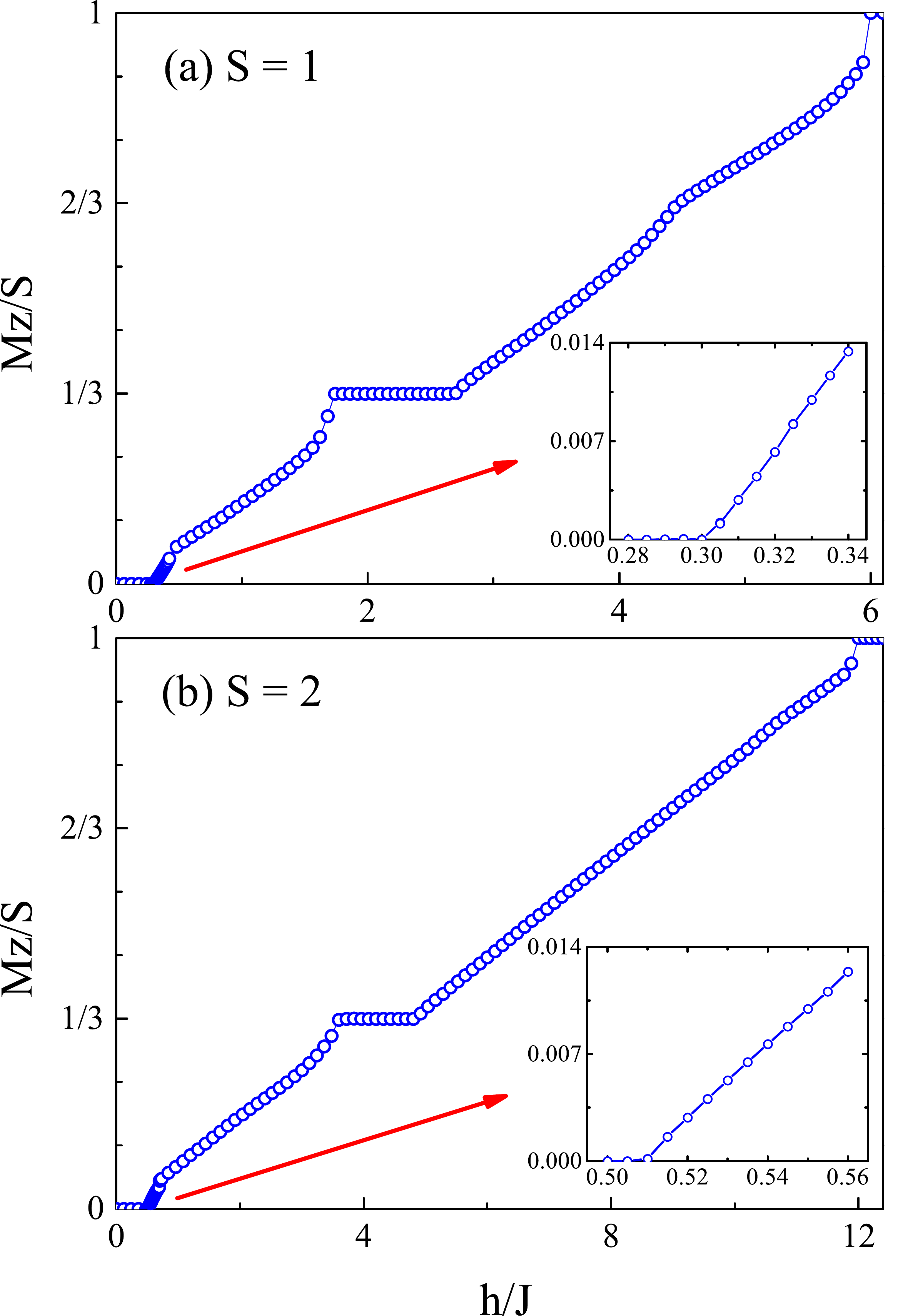,width=7.5cm,angle=0.0}
\end{center}
\caption{(Color online) Longitudinal magnetization, $M_z$, normalized by
its saturation value, $S$, as a function of external magnetic field, $h$,
for (a) $S = 1$ and (b) $S = 2$; calculations performed with bond dimension
$D = 30$. In both cases, zero induced magnetization persists up to a finite
value of the applied field, indicating the presence of a spin gap $\Delta_s
 = h_c \simeq 0.300(5)J$ for the $S = 1$ trimerized simplex solid and $\Delta_s
 = h_c \simeq 0.510(5)J$ for the $S = 2$ uniform simplex solid. Both cases
also show a strong plateau at 1/3 of the saturation magnetization.}
\label{Mh_Spin12}
\end{figure}

For a deeper understanding of the nature of the Heisenberg antiferromagnet on
the Husimi lattice, we have also performed a systematic investigation of the
longitudinal magnetization [Eq.~(\ref{elm})] induced by the application of a
finite magnetic field in Eq.~(\ref{HMF}), as outlined in Sec.~IID3.
Complete results for $S = 1/2$, 3/2, 5/2, 3, 7/2, and 4 are shown in
Fig.~\ref{Mh_tot} and for $S = 1$ and 2 in Fig.~\ref{Mh_Spin12}. All
of these calculations were performed with $D = 20$ and 30, and we find
negligible changes in the results for all cases other than $S = 1/2$ and
$S = 1$ and 2 at small fields. The situation for $S = 1/2$ is already
understood from the results of Fig.~\ref{Ms_Spin0.5}; for $S = 1$ and 2,
some details differ at the percent level at finite magnetizations outside
the simplex-solid state (i.e.~in and just beyond the insets in
Fig.~\ref{Mh_Spin12}). From these observations we conclude that all of our
finite-field calculations are fully representative of the high-$D$ limit.

At low fields, the induced magnetization, $M_z$, reflects directly the
presence or absence of a spin gap. In its presence, no magnetization
can be induced until the applied field exceeds a particular value, breaking
the SU(2) symmetry. For the gapless spin-liquid state at $S = 1/2$, and for
all the N\'eel-ordered ground states shown in Fig.~\ref{Mh_tot}, $M_z$ rises
linearly with $h$ as expected. By contrast, for the simplex-solid ground
states at $S = 1$ and 2 (Fig.~\ref{Mh_Spin12}), zero magnetization is found
to persist until the external field exceeds a critical value, $h_c$, which
corresponds to the spin gap \cite{AMW13}. For the $S = 1$ trimerized
simplex solid we obtain $\Delta_s = h_c \simeq 0.300(5)$ and for the
$S = 2$ uniform simplex solid $\Delta_s = h_c \simeq 0.510(5)$. We find
that the transitions out of the simplex-solid states are continuous (insets,
Fig.~\ref{Mh_Spin12}).

\begin{figure}[t]
\begin{center}
\epsfig{file=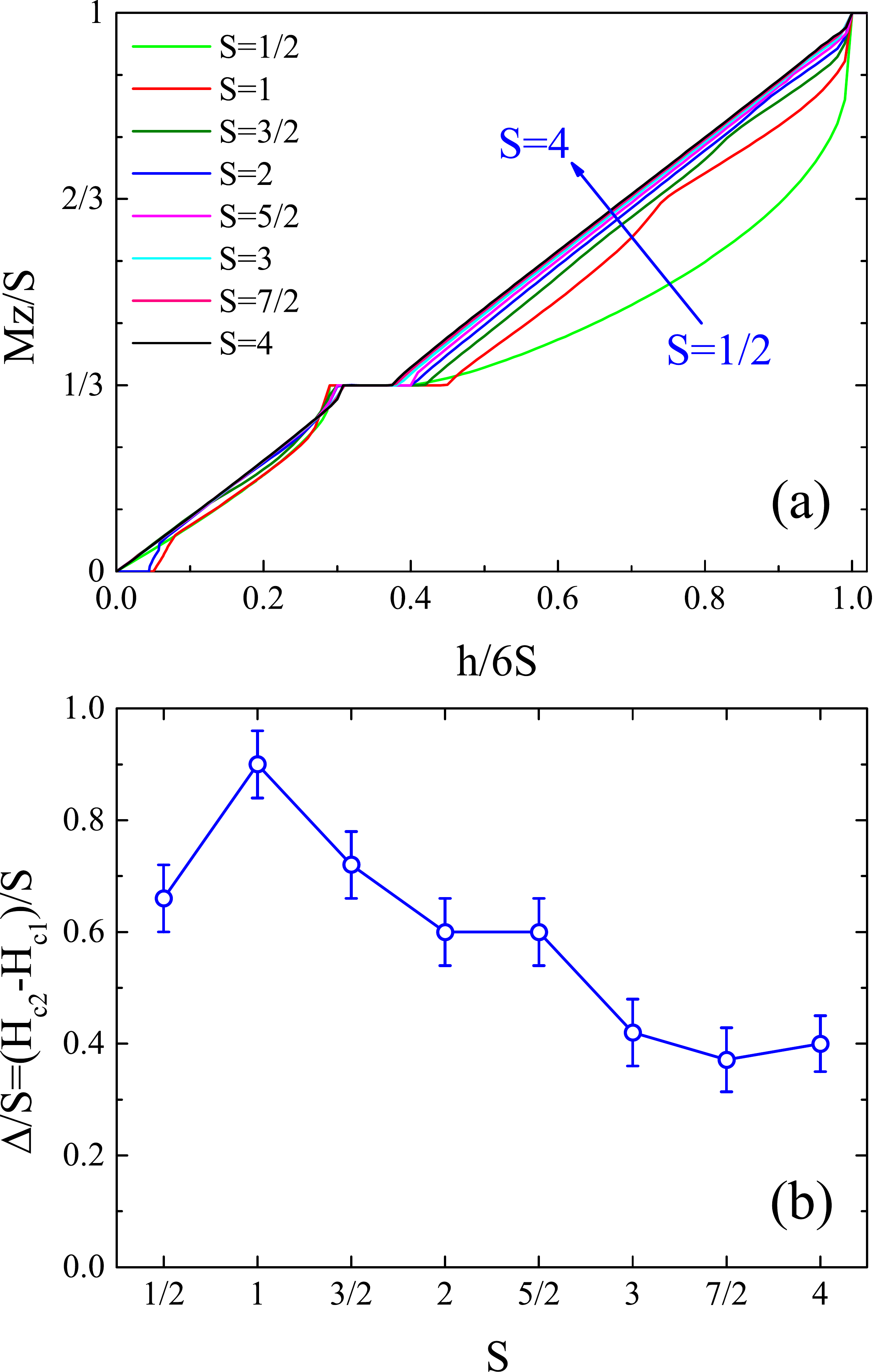,width=7.5cm,angle=0.0}
\end{center}
\caption{(Color online) (a) Comparison of normalized longitudinal magnetization
curves, $M_z(h)$, for all values of the spin quantum number, $S$, computed
with bond dimension $D = 30$. (b) Width of the 1/3-magnetization plateau as
a function of $S$.}
\label{Mh_scaling}
\end{figure}

At finite values of the applied field, the most striking feature is
the presence of a plateau at 1/3 of the saturation magnetizaton. This
plateau is not only present for all values of the spin quantum number, $S$,
both integer and half-odd-integer, but is broad, indicating a robust, gapped
state of the magnetic system around this external field. Before investigating
the nature of the 1/3-plateau state, we comment on the recovery of
classical behavior with increasing $S$. The strongest manifestation of
quantum effects appears to be the curvature in the longitudinal response
above the 1/3 plateau for $S = 1/2$; for all other $S$ values, the
response is significantly more linear, as shown in Fig.~\ref{Mh_scaling}(a).
However, to recover the completely linear magnetization of the classical
limit, it would be necessary for the width of the 1/3 plateau, shown in
Fig.~\ref{Mh_scaling}(b), to vanish. Clearly this situation is not imminent
even at $S = 7/2$ or 4, which reflects again the robust nature of the 1/3
feature.

We comment here that that magnetization curves we have obtained for the
Husimi lattice are quite similar to those obtained for the kagome lattice
in a number of recent studies by exact diagonalization \cite{NS15,CDH+13},
by density matrix renormalization group \cite{NSH13}, and by tensor-network
methods based on infinite PEPS \cite{PZOP15}. These similarities cover all
the primary features of the curves, including spin gap, 1/3 plateau, and
linearity, demonstrating again the extremely close parallels between the
two geometries.

\begin{figure}[t]
\begin{center}
\epsfig{file=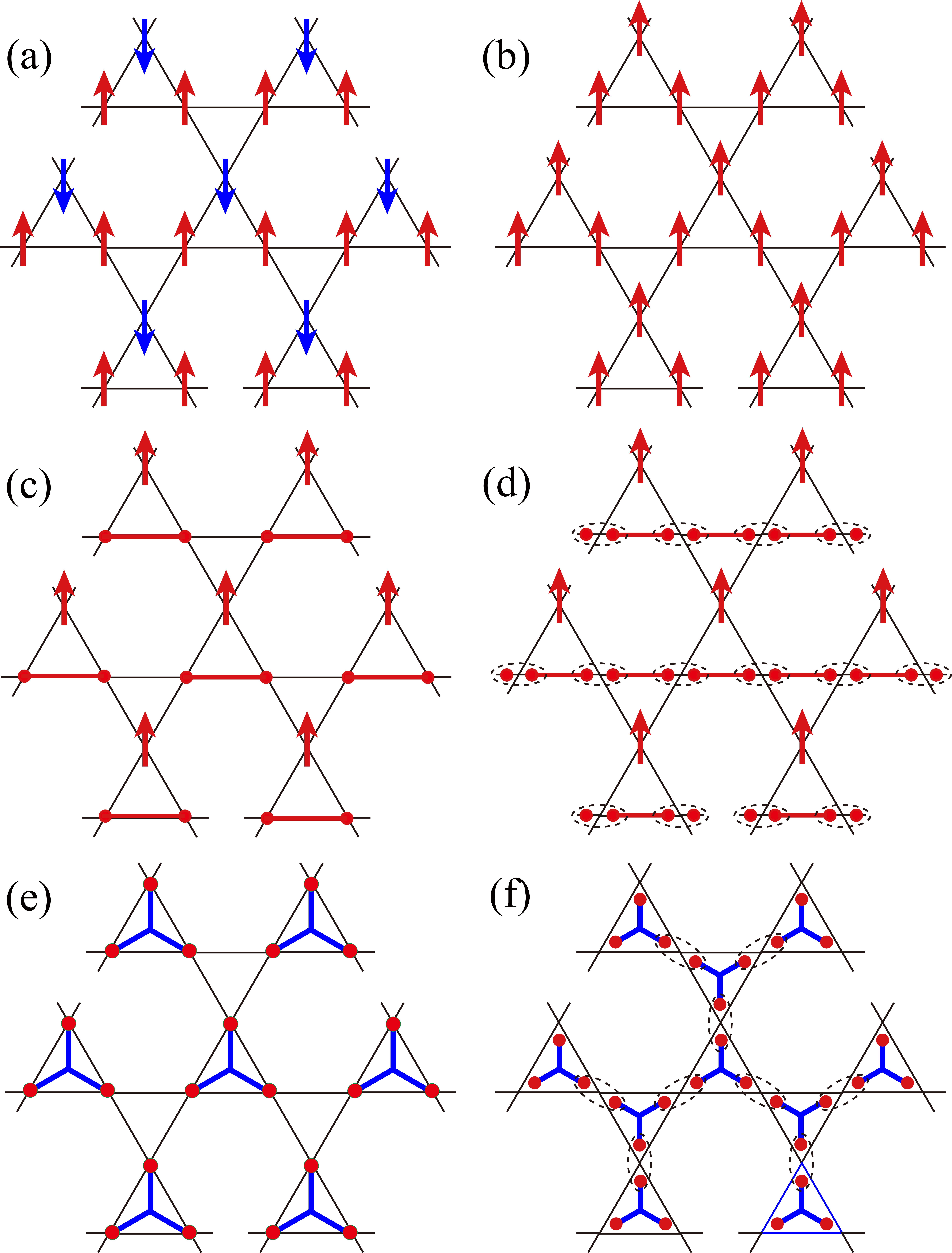,width=8.5cm,angle=0.0}
\end{center}
\caption{(Color online) Six spin states with $m = 1/3$. (a)
``Classical up-up-down'' spin configuration; (b) ``classical ferromagnetic''
spin configuration with site moments $S/3$; (c) ``type-I quantum''
configuration with singlet pairs (red bonds) formed by two physical spins;
(d) ``type-II quantum'' configuration with singlet pairs (red bonds) formed
by two fractionalized virtual spins; (e) non-uniform simplex-solid state with
a triplet on every up-triangle, relevant for $S = 1$; (f) uniform simplex
solid state with a triplet on every simplex, relevant for $S = 2$.}
\label{1-3-plateau}
\end{figure}

\begin{figure*}[t]
\begin{center}
\epsfig{file=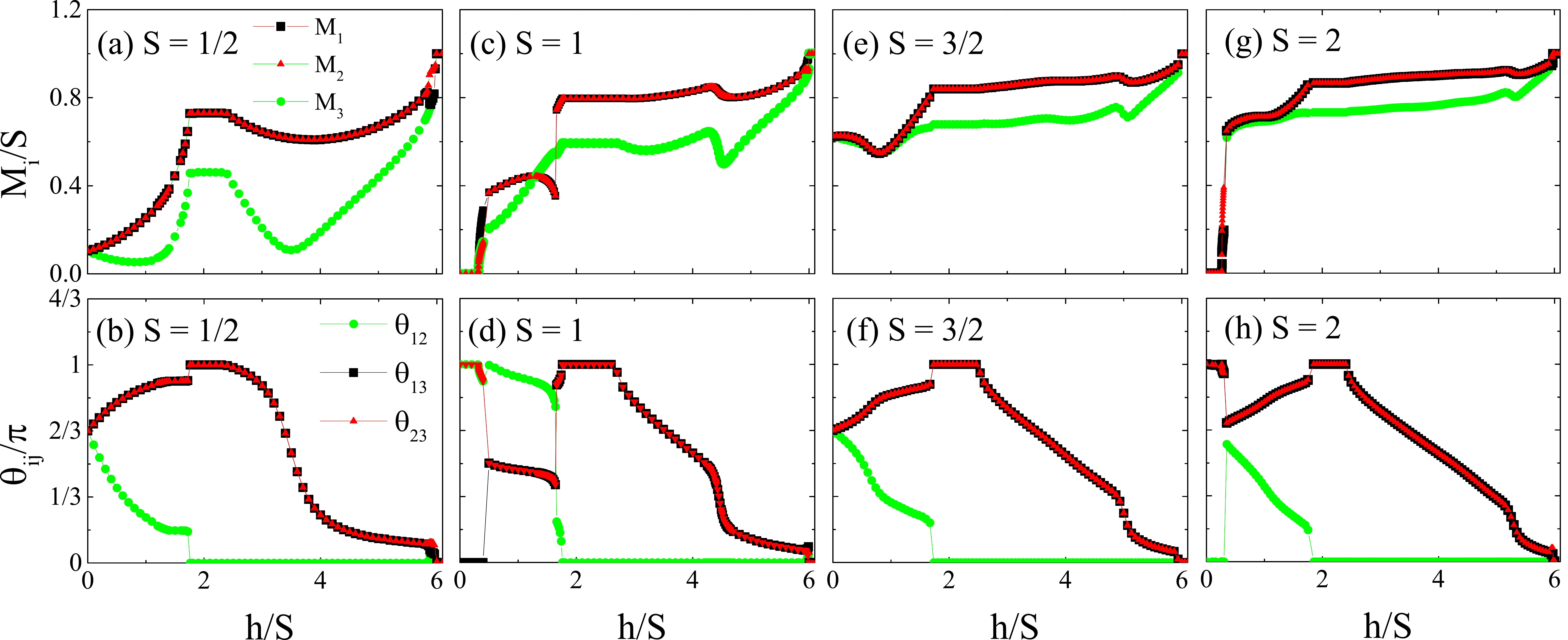,width=17.2cm,angle=0.0}
\end{center}
\caption{(Color online) Magnetic ordering on the Husimi lattice in an applied
field. Shown are the induced ordered moments, $M_i$, along the field ($z$)
axis for each of the three spins on a triangle (a, c, e, g) and the relative
angles, $\theta_{ij}$, between the three pairs of spins on a triangle (b, d,
f, h). The spin quantum numbers are $S = 1/2$ (a, b), $S = 1$ (c, d), $S =
3/2$ (e, f), and $S = 2$ (g, h). Calculations performed with $D = 20$.}
\label{uudc}
\end{figure*}

The primary difference between our results and those of Refs.~\cite{NS15,
PZOP15,CDH+13,NSH13} concerns the presence on the kagome lattice of additional
plateaus in multiples of 1/9 of the saturation magnetization. On the Husimi
lattice, we find no $l/9$ plateaus (where $l$ represents an integer).
Magnetization plateaus generally satisfy the Oshikawa-Yamanaka-Affleck
``commensurate filling'' condition
\cite{Oshikawa03},
\begin{equation}
n (S - m) \;\;\;\; \textrm{ is an integer,}
\label{epc}
\end{equation}
where $n$ denotes the number of sites in the unit cell and $m$ the
longitudinal magnetization. Although we do not observe the plateaus
at $M_z/S = m = (1 - \frac{2}{9S})$ and $(1 - \frac{1}{9S})$ found in
Refs.~\cite{PZOP15,CDH+13,NSH13}, a simple explanation would be that there
are only three sites ($n = 3$) in the unit cell of the $3$-PESS wave
function used in our study. Finding these and higher plateaus would require
that the tensor-network states are based at least on PESS with a 9-site unit
cell, as employed in Ref.~\cite{PZOP15} within a 3-PESS formulation and in
Ref.~\cite{XCY+14} using a 9-PESS. However, the fact that our current
calculations should be able to find a 1/9 plateau for the $S = 3/2$ system
[Eq.~(\ref{epc})] indicates that this is not the only factor involved. In
Refs.~\cite{CDH+13} and \cite{NSH13}, the authors explain the presence of
$l/9$ plateaus on the basis of stable localized eigenstates centered on the
hexagons of the kagome lattice. Thus their absence in the magnetization
response of the Husimi lattice should be understood as a consequence of
the absence of closed loops of triangles.

\subsection{1/3-Plateau State}

We close our analysis of the longitudinal magnetization by investigating the
nature of the 1/3-plateau state for different values of $S$. Our PESS results
contain complete information about the quantum wave function. Many possible
spin configurations could give rise to a net longitudinal magnetization $m =
1/3$, of which six are represented in Fig.~\ref{1-3-plateau}. Some are
effectively magnetically ordered [Figs.~\ref{1-3-plateau}(a) and (b)],
some are motivated by considerations of valence-bond formation
[Figs.~\ref{1-3-plateau}(c) and (d)], and some by simplex solids where the
singlets may be replaced by higher-spin states [Figs.~\ref{1-3-plateau}(e)
and (f)]. These states are not necessarily orthogonal and each is only a
paradigm for the dominant physics of a complex quantum superposition. The
``type-I quantum'' spin configuration [Fig.~\ref{1-3-plateau}(c)], which
contains a singlet pair within a single simplex (triangle), has been
proposed for the $S = 1$ Heisenberg model on the kagome lattice by Cai $et$
$al.$~\cite{CCW09}. Both ``type-I'' and ``type-II quantum'' configurations
may also be relevant to integer-spin simplex-solid states, because their
ground states at zero field contain simplex singlets.

However, analyzing the PESS wave function by calculating the extent of
magnetic order for various $S$ values leads to two possibly unexpected
conclusions. First, it is not necessary to consider the more exotic
singlet- or simplex-based states [Figs.~\ref{1-3-plateau}(c)--(f)]
as candidates for the 1/3 plateau. As shown in Figs.~\ref{uudc}(a), (c),
(e), and (g), the dominant physics of the 1/3-plateau state is a tendency
to magnetic order. This tendency does not occur in the same way in every
case and it is remarkably pronounced for low spins, particularly $S = 1/2$
[Fig.~\ref{uudc}(a)], where the approach to 1/3 magnetization is marked
by a very strong and rapid rise in the ordered component; in fact the loss
of the plateau state at higher fields is accompanied by a decrease in the
overall degree of local spin order before a recovery with some complex
behavior on the very steep approach to full saturation [Fig.~\ref{Mh_tot}(a)].

For $S = 1$ [Fig.~\ref{uudc}(c)], the transition from the gapped,
trimerized simplex-solid state at low fields to a field-ordered state is
abrupt but continuous, although the ordered moment again shows non-monotonic
behavior as the 1/3 and fully aligned states are approached. For $S = 3/2$
[Fig.~\ref{uudc}(e)], the zero-field order of all three spins actually drops
with a small applied field, even as the relative spin angles change, before
increasing again towards the 1/3 plateau. Beyond this, the system tends to
full alignment with only small deviations around $h/S \approx 5$ (possibly
marking an incipient plateau instability, which we cannot trace in the current
3-PESS formalism). For $S = 2$ [Fig.~\ref{uudc}(g)], the ordered moments are
almost homogeneous beyond the continuous transition out of the simplex-solid
phase, and the degree of inhomogeneity and non-monotonic behavior is quite
limited. We comment that for all spins $S \ge 5/2$ (not shown), the evolution
is an increasingly homogeneous and monotonic version of our results for the
N\'eel-ordered $S = 3/2$ state, shown in [Fig.~\ref{uudc}(e)], as the system
approaches the classical limit. Less surprising than all of this complex
behavior in the inequivalent ordered moments is that the ordering configuration
on the 1/3 plateau in all cases is, qualitatively, the up-up-down alignment of
Fig.~\ref{1-3-plateau}(a), as shown in Figs.~\ref{uudc}(b), (d), (f), and (h)
by the evolution of the angles between each the spin pairs on each bond. These
tend to begin near 120-degree angles for all three bonds but evolve towards a
situation with one parallel spin pair before the 1/3 plateau, which then
remains exactly parallel all the way to saturation (Fig.~\ref{uudc}).

\begin{figure}[b]
\begin{center}
\epsfig{file=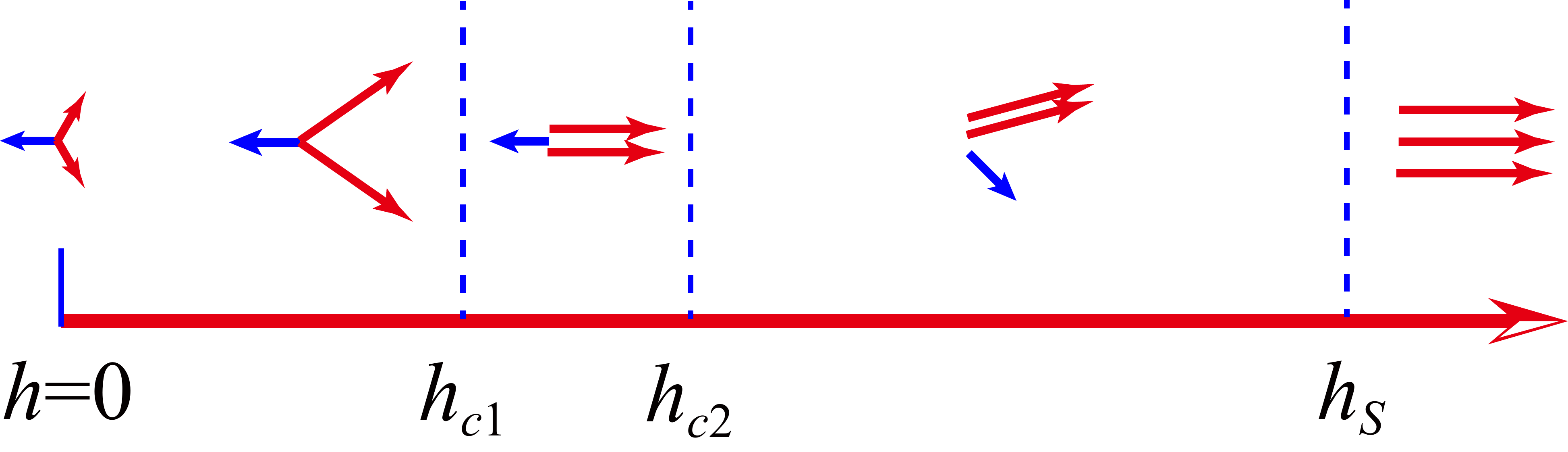,width=8.5cm,angle=0.0}
\end{center}
\caption{(Color online) Schematic representation of the ordered spin
moments and alignments on each triangle across the field-induced phase
diagram for the finite-$D$ $S = 1/2$ Heisenberg model on the Husimi lattice.
The fields $h_{c1}$ and $h_{c2}$ mark the lower and upper boundaries of the
1/3 plateau.}
\label{Mh_pd}
\end{figure}

Secondly, the up-up-down state departs quite significantly from the classical,
rigid-spin form shown in Ref.~\cite{rtm}. There is an immediate and spontaneous
breaking of symmetry at any finite field, or directly beyond the ordering
transition for $S = 1$ and 2, where one of the three spins on each triangle
becomes anti-aligned [Figs.~\ref{uudc}(b), (d), (f), and (h)] and its ordered
moment differs from the other pair. For $S = 1/2$, the average moment of this
third spin actually decreases in size, while the other two grow symmetrically
[Fig.~\ref{uudc}(a)], as represented schematically in Fig.~\ref{Mh_pd}. The
1/3-plateau state realized for every value of $S$ is an asymmetric one,
in which the two spins aligned parallel with the field have a strong ordered
component while the antiparallel spin is weakly ordered; the ordered components
have no universal value, but their sum is 1/3 of the saturation value
corresponding to $S$. On leaving the plateau state, as noted above the
ordered component of the asymmetric spin shrinks in value again for $S = 1$
[Fig.~\ref{uudc}(c)], and very dramatically for $S = 1/2$ [Fig.~\ref{uudc}(a)],
where a reduction is also visible in the symmetric spin moments, indicating
that the magnetic state at higher fields becomes ``more quantum'' again
(i.e.~contains more fluctuations). This behavior does not occur for any
of the higher spins [$S \ge 3/2$, Figs.~\ref{uudc}(e) and (g)], but the
asymmetry in ordered moment between the lone spin and the pair persists
in all cases and for all fields. The bond angles formed by each of the spin
pairs [Figs.~\ref{uudc}(b), (d), (f), and (h)] show a departure again from
full up-up-down field alignment in the regime beyond the 1/3 plateau, but
always with one spin pair completely parallel, as represented in
Fig.~\ref{Mh_pd}. The process of regaining full spin alignment on the
approach to saturation is not a smooth one for any value of $S$, with
rather abrupt changes occurring at high fields, although in contrast to the
ordered moments [Figs.~\ref{uudc}(c), (e), and (g)], it is monotonic in
the bond angles [Figs.~\ref{uudc}(d), (f), and (h)].

We observe that the field-induced magnetization curves show a complex
interplay of a number of quantum fluctuation phenomena. Clearly the spin
configurations in the 1/3-plateau state have significant components of the
``up-up-down'' structure for every value of $S$, even in the most quantum
cases and those most susceptible to simplex-solid formation. However, this
is not precisely the classical configuration of Ref.~\cite{rtm} but an
asymmetric version of it, with one suppressed spin and two extended ones
(Fig.~\ref{Mh_pd}). The degree of up-up-down order is therefore not a
perfectly defined quantity, although a relative degree could certainly
be obtained by comparing only the order of the strong components, or of
the weak ones. We comment that the unequal ordered moments in the graphic
of Fig.~\ref{Mh_pd} are strongly exaggerated for $S = 3/2$ and $S \ge 5/2$,
and would be modified by the simplex-solid (zero-magnetization) states
appearing at low fields for $S = 1$ and $S = 2$. For $S = 1/2$, this
picture is strictly valid only for our results at finite values of $D$,
because the ordered moment at $h = 0$ vanishes in the limit $D \rightarrow
\infty$, but is representative at all finite fields. We expect this schematic
to be accurate in all its details for the phase diagram of the $S = 1/2$
Heisenberg model on the triangular lattice, which does possess 120$^0$-degree
N\'eel order at $h = 0$.

We conclude that the application of a magnetic field is particularly
effective in quenching quantum fluctuation phenomena, driving the system
rapidly to states dominated by magnetic order. However, quantum fluctuation
effects remain very apparent in the clear preference for collinear spin
alignment, manifested in the existence of a robust and well ordered 1/3
plateau for all $S$ values. This effect, which is remarkably strong for
$S = 1/2$, is presumably due to the fully antiferromagnetic quantum
fluctuations allowed between collinear spins. Below collinear configurations
in priority, coplanar ordered spin states remain preferred over non-coplanar
ones.

\section{Summary}

We have investigated the antiferromagnetic Heisenberg model on the triangular
Husimi lattice for values of the quantum spin $S$ up to $S = 4$. We made use
of tensor-network techniques based on Projected Entangled Simplex States
(PESS) to work directly in the thermodynamic limit for highly frustrated
systems. The bisimplex Husimi geometry makes a simple-update approach to
evaluating the quantum wave function almost exact, enabling us to obtain
systematic and highly accurate results to very large values ($D = 260$) of
the bond dimension, and thus to be certain of the trends contained in our
data.

We have demonstrated that the ground state of the model varies widely with
$S$, presenting for $S = 1/2$, 1, 3/2, and 2 excellent examples of four
quite different quantum states. For $S = 1/2$ the ground state is a gapless
(algebraic) spin liquid; for $S = 1$, it is a trimerized (non-uniform)
simplex-solid state with a spin gap; for $S = 3/2$ it is an antiferromagnet
with classical (120-degree) triangular N\'eel order; for $S = 2$, it is a
(uniform) simplex-solid state and therefore is again gapped. However, these
dramatic quantum effects are quenched very rapidily by increasing $S$, and
all higher-spin cases are ordered antiferromagnets, whose ordered moments
increase monotonically with $S$.

One property of a system readily calculated from its tensor-network wave
function is the entanglement spectrum. For the ground states with N\'eel
order, the entanglement spectrum is simple, with all low-lying levels being
non-degenerate, and this result applies also to the gapless spin-liquid state
obtained for $S = 1/2$, which has finite order at all finite values of $D$.
By contrast, the entanglement spectra of the simplex-solid states found
for $S = 1$ and $2$ are clearly different, being characterized by specific
degeneracies in their low-energy levels. Our results suggest that the
entanglement spectrum offers a valuable means of distinguishing between
different types of complex quantum state.

A further quantity readily computed in the PESS framework is the magnetization
response to an external field. We find predictable results at low fields,
with a linear response for the ordered phases and the gapless spin liquid
but a clear spin gap for the simplex-solid states. Surprisingly, however,
no matter how different the low-field quantum states, we find a magnetization
plateau at 1/3 of the saturation value for all values of $S$. This ubiquitous
feature even has the same origin in every case, namely a significant component
of the semiclassical, but asymmetric, ``up-up-down'' configuration on every
triangle. We suggest that the universality of this phenomenon can be traced
to the strong tendency of antiferromagnetic fluctuations to favor collinear
spin alignments.

When considering our results for $S = 1/2$, 1, and 2, it is tempting to seek
a parallel with the Haldane conjecture, that perhaps half-odd-integer-spin
Husimi lattices may be gapless spin liquids whereas integer-spin ones are
simplex solids. However, the rapid emergence of antiferromagnetically ordered
states at (all) higher $S$ values in both series demonstrates that the
predominant physics of the Husimi lattice is two-dimensional. That this
result applies even for a geometry as ``quasi-one-dimensional'' as the
Bethe lattice of triangles allows us to conclude that the dominant effects
on the Husimi lattice are intrinsic to the triangle, and to its local
coordination by only three other triangles, rather than to any features
of the open, tree-like structure.

Persisting with the view of the Husimi lattice as a Bethe lattice of
triangles offers a possible interpretation of our result that the $S = 1/2$
system is a gapless spin liquid. The $S = 1/2$ triangle has two degenerate
doublets and therefore the model may be analogous to a two-color $S = 1/2$
Bethe lattice. The conventional $S = 1/2$ Bethe lattice has been considered
in Ref.~\cite{ro}, where it was found that both the Ising and XY versions
of the model have long-ranged magnetic order, but with orthogonal alignments.
The Heisenberg point therefore appears as the transition between different
magnetic states, which is definitely consistent with our finding of a gapless
spin liquid. This very delicate balance is not reproduced for any other values
of $S$.

Returning to the issue of how weak quantum mechanical fluctuation effects
appear to be on the Husimi lattice, as noted in Secs.~IIIE and IV we have
found that they are restricted to three very small values of $S$ and to
low applied fields. On one hand it is perhaps dispiriting in the search for
exotic quantum states that their phase space is so small even for the Husimi
lattice, which due to its very low coordination and low inter-triangle
connectivity should be a very ``quantum'' geometry. On the other hand,
however, the low connectivity in this case results in relatively low
frustration, restricting it to intra-triangle effects, and this observation
reinforces the fact that the recipe for non-trivial quantum phenomena
requires as essential ingredients both low spin and high frustration.

Finally, we revisit the question of whether our results for the Husimi
lattice shed any important new light on the vexed question of the quantum
ground states of the kagome lattice. All of the properties we have found
for the spin-$S$ Husimi antiferromagnets, including energies, order parameters,
and induced magnetizations, are remarkably similar (both qualitatively and
quantitatively) to those of the kagome lattice where these are known. Clearly
the local structure of corner-sharing triangles determines the vast majority
of the physics, and this is sufficient to cement the parallel in all but the
most delicate cases. Setting aside the higher-spin examples, where the systems
are almost identical \cite{PZOP15}, in this discussion we focus only on
$S = 1$ and $S = 1/2$.

For the $S = 1$ Husimi lattice, our result that the ground state is a
trimerized simplex solid follows mere months after the demonstration, by
two different techniques \cite{CL15,LLW+15}, that the same type of state
has the lowest energy yet obtained for the $S = 1$ kagome lattice. We
suggest that the entanglement spectrum could be used for a definitive
identification of this state. Our results add important new evidence that
such spontaneous breaking of translational symmetry, in the formation of
alternating simplex types, may indeed be the generic physics of the $S = 1$
system.

For the $S = 1/2$ case, our result that the ground state of the Husimi
lattice is a gapless spin liquid requires a more careful interpretation.
The existence of closed loops of triangles in the kagome geometry,
which are absent in the Husimi case, means that geometrical frustration
on the kagome lattice is stronger. Quantum fluctuation effects should
therefore suppress more strongly the magnetic order we find at finite
values of the bond dimension. However, whether this suppression retains
a stronger algebraic form, characteristic of a gapless spin liquid, or
turns over to the exponential form characteristic of a gapped spin liquid,
remains the crucial open question unanswered by the present study.

\section*{Acknowledgments}
We thank W. Li for helpful discussions and J. Richter for valuable comments.
This work was supported by the National Natural Science Foundation of China
(Grant Nos.~10934008, 10874215, and 11174365) and by the National Basic
Research Program of China (Grant Nos.~2012CB921704 and 2011CB309703).

\end{document}